\documentclass[twocolumn,floatfix,10pt]{revtex4-1} \usepackage{amsmath}
\usepackage{amsfonts} \usepackage{amssymb} \usepackage{amsthm}
\usepackage{mathrsfs} \usepackage{graphicx} \usepackage{txfonts}
\usepackage{multirow} \usepackage{color} \usepackage[caption=false]{subfig}
\usepackage{array} \usepackage{enumitem}

\usepackage[colorlinks, urlcolor=black, citecolor=blue, linkcolor=black]{hyperref}

\def \Re{\text{Re}} \def \Im{\text{Im}} \def \Br{\text{Br}}

\def\nn {\nonumber} \def\sss{\scriptscriptstyle} 

\def\gev{\ensuremath{\mathrm{Ge\kern -0.1em V}}}
\def\tev{\ensuremath{\mathrm{Te\kern -0.1em V}}}

\def\gammafJ{\ensuremath{\mathrm{\Gamma_{\!f}^{\text{\fontsize{0.2cm}{
            1em}\selectfont(J)}}}~}}
\def\gammaJ{\ensuremath{\mathrm{\Gamma^{\!\text{\fontsize{0.2cm}{1em}
          \selectfont(J)}}}~}}

\def\barp{{\raise.35ex\hbox{${\sss (}$}}---{\raise.35ex \hbox{${\sss
)}$}}}
\def\bdbarp{\hbox{$B_d$\kern-1.4em\raise1.4ex\hbox{\barp}}}
\def\Dbarp{\hbox{$D$\kern-.85em\raise1.2ex\hbox{{{\raise.35ex
\hbox{{\tiny (}}}--{\raise.35ex\hbox{${\sss )}$}}}}}}
\def\deltabarp{\hbox{$\delta_0$\kern-1.08em\raise1.3ex
\hbox{{{\raise.35ex\hbox{{\tiny (}}}--{\raise.35ex\hbox{${\sss
)}$}}}}}}

\newcommand{\ket}[1]{\left| #1 \right\rangle}

\newcommand{\modulus}[1]{\left| #1 \right|}

\newcommand{\even}[2]{e_{\text{\fontsize{0.2cm}{1em}\selectfont #1}
}^{\text{\fontsize{0.2cm}{1em}\selectfont (#2)}}}
\newcommand{\odd}[2]{o_{\text{\fontsize{0.2cm}{1em}\selectfont #1}
}^{\text{\fontsize{0.2cm}{1em}\selectfont (#2)}}}
\newcommand{\Even}[2]{E_{\text{\fontsize{0.2cm}{1em}\selectfont #1}
}^{\text{\fontsize{0.2cm}{1em}\selectfont (#2)}}}
\newcommand{\Odd}[2]{O_{\text{\fontsize{0.2cm}{1em}\selectfont #1}
}^{\text{\fontsize{0.2cm}{1em}\selectfont (#2)}}}
\newcommand{\Ampe}[2]{A_{\text{\fontsize{0.2cm}{1em}\selectfont E
\!}#1}^{\text{\fontsize{0.2cm}{1em}\selectfont (#2)}}}
\newcommand{\Ampo}[2]{A_{\text{\fontsize{0.2cm}{1em}\selectfont O
\!}#1}^{\text{\fontsize{0.2cm}{1em}\selectfont (#2)}}}

\newcommand{\Fe}[2]{F_{\!\text{\fontsize{0.2cm}{1em}\selectfont E
\!}#1}^{\text{\fontsize{0.2cm}{1em}\selectfont (#2)}}}
\newcommand{\Fo}[2]{F_{\!\text{\fontsize{0.2cm}{1em}\selectfont O
\!}#1}^{\text{\fontsize{0.2cm}{1em}\selectfont (#2)}}}
\newcommand{\Fest}[2]{F_{\!\text{\fontsize{0.2cm}{1em}\selectfont E
\!}#1}^{\text{\fontsize{0.2cm}{1em}\selectfont (#2)}*}}
\newcommand{\Fost}[2]{F_{\!\text{\fontsize{0.2cm}{1em}\selectfont O
\!}#1}^{\text{\fontsize{0.2cm}{1em}\selectfont (#2)}*}}

\newcommand{\Tone}[1]{T_{\!\text{\fontsize{0.2cm}{1em}\selectfont
1}}^{ \text{\fontsize{0.2cm}{1em}\selectfont (#1)}}}
\newcommand{\Ttwo}[1]{T_{\!\text{\fontsize{0.2cm}{1em}\selectfont
2}}^{ \text{\fontsize{0.2cm}{1em}\selectfont (#1)}}}
\newcommand{\Tonep}[1]{T_{\!\text{\fontsize{0.2cm}{1em}\selectfont
1}}^{ \prime \text{\fontsize{0.2cm}{1em}\selectfont (#1)}}}
\newcommand{\Ttwop}[1]{T_{\!\text{\fontsize{0.2cm}{1em}\selectfont
2}}^{ \prime \text{\fontsize{0.2cm}{1em}\selectfont (#1)}}}
\newcommand{\Ti}[1]{T_{\!\text{\fontsize{0.2cm}{1em}\selectfont
i}}^{ \text{\fontsize{0.2cm}{1em}\selectfont (#1)}}}
\newcommand{\Tip}[1]{T_{\!\text{\fontsize{0.2cm}{1em}\selectfont
i}}^{ \prime \text{\fontsize{0.2cm}{1em}\selectfont (#1)}}}

\newcommand{\Uone}[1]{U_{\!\text{\fontsize{0.2cm}{1em}\selectfont
1}}^{ \text{\fontsize{0.2cm}{1em}\selectfont (#1)}}}
\newcommand{\Utwo}[1]{U_{\!\text{\fontsize{0.2cm}{1em}\selectfont
2}}^{ \text{\fontsize{0.2cm}{1em}\selectfont (#1)}}}
\newcommand{\Ui}[1]{U_{\!\text{\fontsize{0.2cm}{1em}\selectfont
i}}^{ \text{\fontsize{0.2cm}{1em}\selectfont (#1)}}}

\newcommand{\Vone}[1]{V_{\!\text{\fontsize{0.2cm}{1em}\selectfont
1}}^{ \text{\fontsize{0.2cm}{1em}\selectfont (#1)}}}
\newcommand{\Vtwo}[1]{V_{\!\text{\fontsize{0.2cm}{1em}\selectfont
2}}^{ \text{\fontsize{0.2cm}{1em}\selectfont (#1)}}}
\newcommand{\Vi}[1]{V_{\!\text{\fontsize{0.2cm}{1em}\selectfont
i}}^{ \text{\fontsize{0.2cm}{1em}\selectfont (#1)}}}

\newcommand{\Xz}{\ensuremath{X^{\text{\fontsize{0.2cm}{1em}\selectfont
(0)}}}}
\newcommand{\Xo}{\ensuremath{X^{\text{\fontsize{0.2cm}{1em}\selectfont
(1)}}}}
\newcommand{\Xt}{\ensuremath{X^{\text{\fontsize{0.2cm}{1em}\selectfont
(2)}}}}
\newcommand{\Xot}{\ensuremath{\tilde{X}^{\text{\fontsize{0.2cm}{1em}
\selectfont
(1)}}}}

\newcommand{\Lagr}[1]{\mathscr{L}^{\!\text{\fontsize{0.2cm}{1em}
\selectfont (#1)}} _{\text{\fontsize{0.2cm}{1em}\selectfont XZZ}}}

\newcommand{\Vertex}[1]{V^{#1}_{\text{\fontsize{0.2cm}{1em}\selectfont
XZZ}}}

\newcommand{\Delt}[1]{\Delta^{\!\text{\fontsize{0.2cm}{1em}\selectfont
(#1)}}}

\def\NJ{\ensuremath{\mathscr{N}^{\!\text{\fontsize{0.2cm}{1em}
\selectfont (J)}}}}
\def\SJ{\ensuremath{S^{\!\text{\fontsize{0.2cm}{1em}\selectfont
(J)}}}}

\allowdisplaybreaks

\begin{document}

\title{Disentangling the Spin-Parity of a Resonance via the Gold-Plated Decay
Mode}
  
\author{Tanmoy Modak} \email[E-mail at: ]{tanmoyy@imsc.res.in} \affiliation{The
Institute of Mathematical Sciences, Taramani, Chennai 600113, India}

\author{Dibyakrupa Sahoo} \email[E-mail at: ]{sdibyakrupa@imsc.res.in}
\affiliation{The Institute of Mathematical Sciences, Taramani, Chennai 600113,
India}

\author{Rahul Sinha} \email[E-mail at: ]{sinha@imsc.res.in} \affiliation{The
Institute of Mathematical Sciences, Taramani, Chennai 600113, India}

\author{Hai-Yang Cheng} \email[E-mail at: ]{phcheng@phys.sinica.edu.tw}
\affiliation{Institute of Physics, Academia Sinica, Taipei, Taiwan 11529,
Republic of China}

\author{Tzu-Chiang Yuan} \email[E-mail at: ]{tcyuan@phys.sinica.edu.tw}
\affiliation{Institute of Physics, Academia Sinica, Taipei, Taiwan 11529,
Republic of China}
  
\date{\today}

\begin{abstract}
  Searching for new resonances and finding out their properties is an essential
  part of any existing or future particle physics experiment. The nature of a
  new resonance is characterized by its spin, charge conjugation, parity, and
  its couplings with the existing particles of the Standard Model.  If a new
  resonance is found in the four lepton final state produced via two
  intermediate $Z$ bosons, the resonance could be a new heavy scalar or a $Z'$
  boson or even a higher spin particle. In such cases the step by step
  methodology as enunciated in this paper can be followed to determine the spin,
  parity and the coupling to two $Z$ bosons of the parent particles, in a fully
  model-independent way. In our approach we show how three uniangular
  distributions and few experimentally measurable observables can conclusively
  tell us about the spin, parity as well as the couplings of the new resonance
  to two $Z$ bosons. We have performed a numerical analysis to validate our
  approach and showed how the uniangular observables can be used to disentangle
  the spin parity as well as coupling of the resonance.
\end{abstract}

\maketitle

\section{Introduction}\label{sec:intro}

With the recent discovery of the `Higgs' boson
\cite{Aad:2012tfa,Chatrchyan:2012ufa}, all the ingredients of the standard model
of particle physics (SM) have been found. However, we do know that the SM does
not fully explain the whole of the nature at its most fundamental level. For
example, the problem of naturalness, the existence of extremely small masses for
the neutrinos required to explain the observed neutrino oscillations, the
abundance of matter over anti-matter in our observable universe, the
constituents of dark matter (which is about five times more abundant than the
ordinary matter) are a few of many issues which can not be handled in SM. So SM
encompasses an incomplete description of nature and hence it must be
supplemented or extended with some other hitherto unknown new physics. Any model
of new physics invariably includes new interactions and thus many new particles.
In order to have a comprehensive view of new physics it is therefore essential
to look for new fundamental particles in experiments such as the Large Hadron
Collider (LHC) or in the proposed future experiments such as the International
Linear Collider (ILC). It is however important to have model independent methods
at place to characterize resonances that would be observed in these high
luminosity and high energy experiments. This paper is a step forward in that
direction.

Similar idea was espoused in Ref.~\cite{Modak:2013sb} in connection with the
$125$-$126~\gev$ `Higgs' resonance. Since the `Higgs' was found to decay to two
photons, the spin-1 possibility in this case was ruled out using Landau-Yang
theorem. This paper, however, has the complete theoretical analysis and we do
consider the spin-1 possibility here. The decay of a resonance to four leptons
via a pair of $Z$ bosons is considered as the gold plated mode for discovery of
any resonance in a hadron collider such as LHC. This mode can be used to search
for many new resonances, such as a heavy scalar resonance, a $Z'$ boson or a
Kaluza-Klein boson or a massive graviton. Hence, all the spin possibilities must
be considered. We analyse the decay of a resonance (say, $X$ with spin $J$) via
this golden channel: $X \to Z^{(*)} Z^{(*)} \to (\ell_1^- \ell_1^+) (\ell_2^-
\ell_2^+)$, where $\ell_1$, $\ell_2$ are leptons $e$ or $\mu$. Since we are
considering an unknown particle with unknown mass, it may be heavy enough to
produce two real $Z$ bosons or we can have one on-shell $Z$ and another
off-shell $Z$, or both the $Z$'s can be off-shell. We emphasize that the final
state $(e^+e^-)(\mu^+\mu^-)$ is not equivalent to $(e^+e^-)(e^+e^-)$ or
$(\mu^+\mu^-)(\mu^+\mu^-)$ as sometimes mentioned in the literature, since the
latter final states have to be anti-symmetrized with respect to each of the two
sets of identical fermions in the final state. The anti-symmetrization of the
amplitudes is not done in our analysis and hence our analysis applies only to
$(e^+e^-)(\mu^+\mu^-)$. Since $X$ decays to two $Z$ bosons which are vector
bosons, $X$ can have spin possibilities $J=0,1,2$. Higher spin possibilities ($J
\geqslant 3$) need not be considered as the number of independent helicity
amplitudes remains three (for higher odd integer spins) or six (for higher even
integer spins), and the only change in the amplitude comes from extra powers of
momentum of the $Z$ bosons~\cite{ref:Higher-spin}. This was shown by an example
in Ref.~\cite{Modak:2013sb}. In this paper we examine the angular distributions
for the different spins $J = 0,1,2$ and parity of such a new resonance $X$ and
present a strategy to determine them, as well as the couplings of $X$ to the
pair of $Z$ bosons with the help of a few well defined experimental observables.

We start by considering the most general effective Lagrangian for each spin
possibility of $X$ decaying to two $Z$ bosons and then write down the
corresponding decay vertices. We evaluate the partial decay rate of $X$ in terms
of the invariant mass squared of the dileptons produced from the two $Z$ decays
and the angular distributions of the four lepton final state. We demonstrate
that by studying three uni-angular distributions (i.e.\ \textit{distributions
involving only one angle}) one can almost completely determine the spin and
parity of $X$ and also explore any anomalous couplings in the most general
fashion. We provide some experimental observables that can be tested to predict
whether the resonance is a parity eigenstate or not irrespective of its spin.
Then we analyse the nuances of differences in the uni-angular distributions
which we take into consideration for separating each of the spin, parity
possibilities. For this we express our uni-angular distributions in terms of
helicity amplitudes in the transversity basis which are very effective in their
sensitivity to the parity of the parent particle. We find that the $J=1$
possibility can be easily distinguished from the $J=0,2$ possibilities since the
uni-angular distributions for $J=1$ are completely predictable.

Nelson~\cite{Nelson:1984bb, Dell'Aquila:1985ve, Nelson:1986ki} and
Dell'Aquilla~\cite{Dell'Aquila:1985ve} realized the significance of studying
angular correlations in such processes as Higgs ($J=0$) boson decaying to a pair
of $Z$ bosons for inferring the nature of the Higgs boson.
Refs.~\cite{Miller:2001bi, Choi:2002dq, Choi:2002jk, Gao:2010qx} were first to
extend the analysis to include higher spin possibilities.  We study similar
angular correlations in this paper to unambiguously determine the spin and
parity of any new resonance decaying to four leptons via two $Z$ boson.  We
begin the study by considering the most general $XZZ$ vertices for $J=0$, $J=1$
and $J=2$ resonance $X$.  Then we find out the effective Lagrangians and
uniangular distributions in terms of different observables for each spin
possibilities and layout a procedure to identify them. We emphasize that the
full angular distribution for the decay is described in terms of orthogonal
function of sine and cosine of the angles involved in the decay. Hence, no
information is lost by considering uniangular distributions.

Decay of a resonance into two $Z$ bosons which then decay into four leptons has
been well studied in literature for different spin possibilities.  For example
Refs.~\cite{Kramer:1993jn, Buszello:2002uu, Bluj:2006, Eboli:2011bq,
Stolarski:2012ps, Bolognesi:2012mm, Han:1998sg, Gainer:2014hha, Keung:2008ve,
Langacker:2008yv,Flacke:2012ke} as well as Refs.~\cite{Miller:2001bi,
Choi:2002dq, Choi:2002jk, Gao:2010qx} discuss extensively coupling of $X$ to two
$Z$ bosons for different spin possibilities as well as how to determine the
spin-parity of a new resonance decaying into four leptons via a pair of $Z$
bosons. More recent references for Higgs decay to four leptons via two $Z$
bosons can be found in Refs.~\cite{Modak:2013sb, Gainer:2014hha, Avery:2012um,
Gainer:2013rxa, Chen:2013waa}.  The aim of the current paper should not be
confused with that of Ref.~\cite{Modak:2013sb}, which dealt only with the
$125~\gev$ Higgs case. The current paper significantly improves upon the
concepts presented in Ref.~\cite{Modak:2013sb} by \textit{generalizing them to
search for new resonances of arbitrary mass and arbitrary spin-parity using the
gold-plated decay mode}.  The analysis now includes the possibility of a spin-1
resonance which can also decay to two on-shell $Z$ bosons, a case which was not
dealt with earlier.  Ref.~\cite{Bolognesi:2012mm} does consider the spin-1
possibility. However, we emphasize that our analysis differs from this analysis
as well, in many significant aspects. Our analysis does not depend on the
resonance production mechanism. Unlike Ref.~\cite{Bolognesi:2012mm}, we start
with the most general effective Lagrangian for any spin and parity possibility.
Our treatment in terms of transversity amplitudes is explicit on their parity
behavior whereas the amplitudes of the other reference are ambivalent. We
carefully use Schouten identity to show which specific combination of vertex
factors contribute to the orthogonal transversity amplitudes.

We show numerically that the uniangular distributions can indeed be used to
disentangle the spin and parity of the resonance $X$. We benchmark our analysis
for the $J=1$ possibility of $X$ and calculate the values of the angular
asymmetries for the $14~\tev$ and $33~\tev$ LHC.  The analysis can also be
extended for $J=2$ spin possibility.

\begin{figure}[hbtp]
\centering %
\includegraphics[scale=0.9]{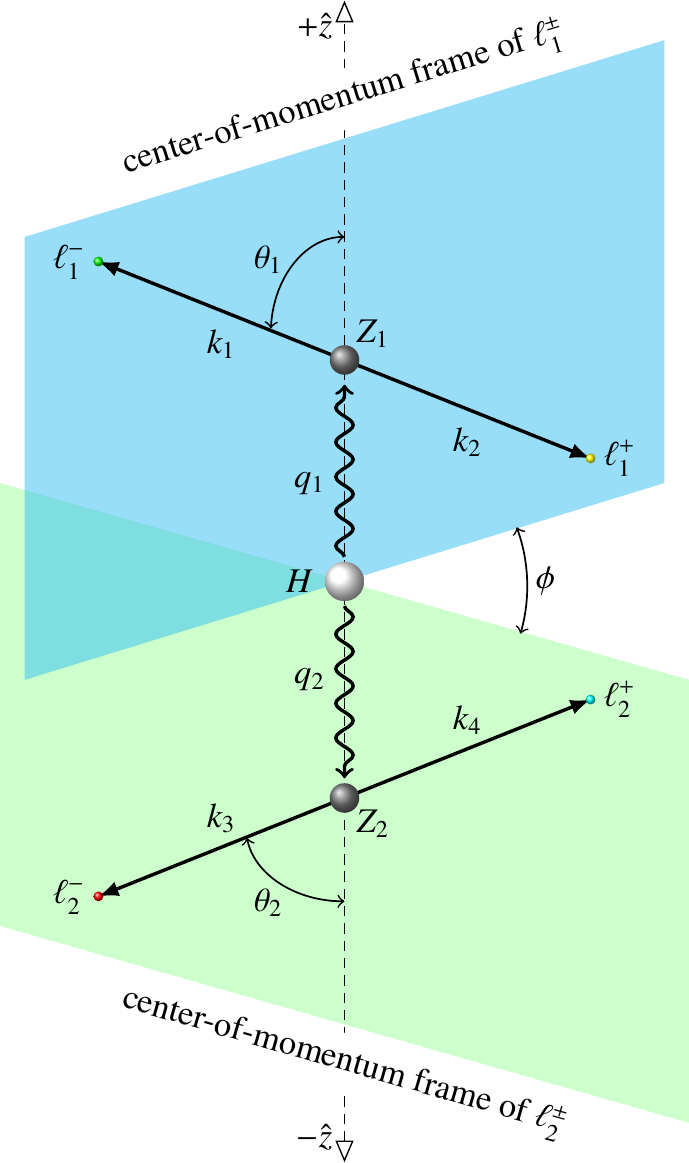} %
\caption{Definition of the polar angles ($\theta_1$ and $\theta_2$) and the
azimuthal angle ($\phi$) in the decay of $X$ to a pair of $Z$'s, and then to
four charged leptons: $X\to Z_1 + Z_2 \to (\ell_1^- + \ell_1^+) + (\ell_2^- +
\ell_2^+),$ where $\ell_1, \ell_2 \in \{ e,\mu \}$. It should be clear from the
figure that $\vec{k}_1 = -\vec{k}_2$ and $\vec{k}_3 = -\vec{k}_4$.  The lepton
pairs are reconstructed in their respective center-of-momentum frames. The angle
between the two decay planes is the angle $\phi$. The spin projection of
particle $X$ is along the $\hat{z}$ direction.} %
\label{fig:XZ1Z2}
\end{figure}

A few words about the notation followed in this paper is noteworthy. Since our
aim in this paper is to get an insight into the spin and parity of the resonance
under consideration, the notation has been carefully designed in such a way that
at any stage in the paper one would have no difficulty in recognizing the terms
(in Lagrangian or vertex factor or amplitude) which contribute for even and odd
parity cases. We use $\even{i}{J}$ ($\odd{i}{J}$), $\Even{i}{J}$ ($\Odd{i}{J}$),
and $\Ampe{i}{J}$ ($\Ampo{i}{J}$) for parity-even (parity-odd) coupling
constants, form factors and helicity amplitudes respectively for spin $J$ case
in general. All the notations are described when they are introduced in the
paper.

Sec.~\ref{sec:analysis} deals completely with the stepwise building of the
theoretical analysis. In subsection~\ref{subsec:Lagrangian-vertex} we write down
most general Lagrangian and corresponding vertex factor for each possible spin.
Transversity amplitudes and uni-angular distributions for each of the spin
possibilities are analyzed in subsections~\ref{subsec:transversity-amplitudes}
and \ref{subsec:uniangular} respectively.  In Sec.~\ref{sec:numerical} we
discuss about the possibility of finding one such resonance and investigate how
to disentangle them. In subsection~\ref{subsec:1p} we find the uniangular
distributions and the values of the observables for a spin $1^{+}$ heavy
resonance in future $14~\tev$ and $33~\tev$ LHC run. We repeat the analysis for
a spin $1^{-}$ resonance in subsection~\ref{subsec:1m}.  We summarize in
Sec.~\ref{sec:conclusion} emphasizing the importance of our results.

\section{Analysis of the decay of a resonance to four final charged leptons via two \texorpdfstring{$Z$}{Z} Bosons} \label{sec:analysis}

Let us consider the decay of $X$ to four charged leptons via a pair of
$Z$ bosons: $$X \to Z_1 + Z_2 \to (\ell_1^- + \ell_1^+) + (\ell_2^- +
\ell_2^+),$$ where $\ell_1$, $\ell_2$ are leptons $e$ or $\mu$. As
mentioned in the introduction we assume $\ell_1$ and $\ell_2$ are not
identical. The kinematics for the decay is as shown in
Fig.~\ref{fig:XZ1Z2}. The resonance $X$ at rest decays to two $Z$
bosons (none of them or either of them or both of them can be on-shell
depending on the mass of $X$) $Z_1$ and $Z_2$ moving along the
$+\hat{z}$ and $-\hat{z}$ directions with four-momenta $q_1$ and $q_2$
respectively. The decays of $Z_1$ and $Z_2$ are considered in their
respective rest frames. The angles and momenta involved are as
described in Fig.~\ref{fig:XZ1Z2}. The four-momentum of $X$ is defined
as $P$: $P = q_1 + q_2$. We also define $Q = q_1 - q_2$. We choose
$Z_1$ to decay to lepton pair $\ell_1^\pm$ with momenta $k_1$ and
$k_2$ respectively and $Z_2$ to decay to $\ell_2^\pm$ with momenta
$k_3$ and $k_4$ respectively. We define two unit normals $\vec{n}_1$
and $\vec{n}_2$ to the planes containing $\vec{k}_1, \vec{k}_2$ and
$\vec{k}_3, \vec{k}_4$ respectively by
\begin{align}
\vec{n}_1 &= \frac{\vec{k}_1 \times \vec{k}_2}{\modulus{\vec{k}_1 \times
\vec{k}_2}} = - \hat{y},\\ %
\vec{n}_2 &= \frac{\vec{k}_3 \times \vec{k}_4}{\modulus{\vec{k}_3 \times
\vec{k}_4}} = - \left( \sin\phi \; \hat{x} - \cos\phi \; \hat{y} \right).
\end{align}
Thus the azimuthal angle $\phi$ can be specified unambiguously by
\begin{align}
\vec{n}_1 \cdot \vec{n}_2 &= -\cos\phi,\\ %
\vec{n}_1 \times \vec{n}_2 &= -\sin\phi \; \hat{z}.
\end{align}
We begin the study by considering the most general $XZZ$ interaction Lagrangians
for a $J=0$ , a $J=1$ and a $J=2$ resonance $X$. From the Lagrangians we find
out the most general $XZZ$ decay vertex factor for each spin possibility.

\subsection{Most general interaction Lagrangian and Vertex factor for the decay \texorpdfstring{$X \to ZZ$}{X->ZZ}} \label{subsec:Lagrangian-vertex}

Considering Lorentz invariance one can write down the most general interaction
Lagrangian for the decay of $X$ to two $Z$ bosons. The Lagrangian depends on the
spin of the parent resonance $X$. We denote the most general Lagrangian for
spin-$J$ resonance $X$ decaying to two $Z$ bosons by $\Lagr{J}$. The Lagrangians
for the allowed spin possibilities are given by
\begin{itemize}[leftmargin=4mm]
\item[$\bullet$] $J=0$:
\begin{align}
\Lagr{0} =& \even{1}{0} \; \Xz \; Z^{\alpha} \; Z^{\beta} \; g_{\alpha\beta} +
\even{2}{0} \; \Xz \; Z^{\mu\nu} \; Z_{\mu\nu} \nn\\ %
&+ i \; \odd{1}{0} \; \Xz \; \tilde{Z}^{\mu\nu} \; Z_{\mu\nu},
\label{eq:Lagrangian-0}
\end{align}
\item[$\bullet$] $J=1$:
\begin{align}
\Lagr{1} =& \odd{1}{1} \; \Xo_{\mu} \; Z^{\mu\nu} \; Z_{\nu} + \odd{2}{1} \;
\Xo_{\mu\nu} \; Z^{\mu} \; Z^{\nu} \nn\\ %
&  + i \; \even{1}{1} \; \Xo_{\mu} \; \tilde{Z}_{\mu\nu} \; Z^{\nu} + i \;
\even{2}{1} \; \Xot_{\mu\nu} \; Z^{\mu} \; Z^{\nu}, \label{eq:Lagrangian-1}
\end{align}
\item[$\bullet$] $J=2$:
\begin{align}
\Lagr{2} =& \even{1}{2} \; \Xt_{\mu\nu} \, Z^{\mu} \, Z^{\nu} + \even{2}{2} \;
\Xt_{\mu\nu} \, Z^{\mu\alpha} \, Z^{\nu\beta} \, g_{\alpha\beta} \nn \\ %
& + \even{3}{2}  \Xt_{\mu\nu} \, \left( \partial_{\beta} \, Z^{\mu\alpha}
\right) \left( \partial_{\alpha} \, Z^{\nu\beta} \right) \nn\\ %
&+ \even{4}{2} \left( \partial_{\alpha} \, \partial_{\beta} \, \Xt_{\mu\nu}
\right) \, \left( Z^{\mu\alpha} Z^{\nu\beta} \right) \nn \\ %
& + \even{5}{2} \Xt_{\mu\nu} \left( \partial^{\mu} \, Z^{\alpha\beta} \right) \,
\left( \partial^{\nu} \, Z_{\alpha\beta} \right) \nn\\ %
& + i \; \odd{1}{2} \; \Xt_{\mu\nu} \, \tilde{Z}^{\mu\alpha} \, Z^{\nu\beta} \,
g_{\alpha\beta} \nn \\ %
& + i \; \odd{2}{2} \; \Xt_{\mu\nu} \, \left( \partial_{\beta} \,
\tilde{Z}^{\mu\alpha} \right) \left( \partial_{\alpha} \, Z^{\nu\beta} \right)
\nn\\ %
& + i \; \odd{3}{2}  \left( \partial_{\alpha} \, \partial_{\beta} \,
\Xt_{\mu\nu} \right) \, \left( \tilde{Z}^{\mu\alpha} Z^{\nu\beta} \right) \nn \\
& + i \; \odd{4}{2} \; \Xt_{\mu\nu} \, \left( \partial^{\mu} \,
\tilde{Z}^{\alpha\beta} \right) \, \left( \partial^{\nu} \, Z_{\alpha\beta}
\right), \label{eq:Lagrangian-2}
\end{align}
\end{itemize}
where $\even{i}{J}$ and $\odd{i}{J}$ are the effective coupling constants for a
specified spin $J$ of the parent particle that come with the parity-even and
parity-odd parts of the Lagrangian respectively; $\Xz, \Xo_{\mu}, \Xt_{\mu\nu}$
are the scalar, vector and tensor fields for the corresponding spin
possibilities of the particle $X$; the tensor $\Xt_{\mu\nu}$ is a symmetric,
traceless and divergenceless tensor; $Z^{\alpha}$ is the vector field for the
$Z$ boson; and $\Xo_{\mu\nu}$, $\Xot_{\mu\nu}$, $Z_{\mu\nu}$,
$\tilde{Z}_{\mu\nu}$ are defined as
\begin{align}
\Xo_{\mu\nu} &= \partial_{\mu} \Xo_{\nu} - \partial_{\nu} \Xo_{\mu},\\ %
\Xot_{\mu\nu} &= \epsilon_{\mu\nu\rho\sigma} \,
X^{\text{\fontsize{0.2cm}{1em}\selectfont (1)}\rho\sigma},\\ %
Z_{\mu\nu} &= \partial_{\mu} Z_{\nu} - \partial_{\nu} Z_{\mu},\\ %
\tilde{Z}_{\mu\nu} &= \epsilon_{\mu\nu\rho\sigma} \, Z^{\rho\sigma}.
\end{align}
It is noteworthy to clearly emphasize that the two quantities $\Xo_{\mu\nu}$ and
$\Xt_{\mu\nu}$ must not be confused. The quantity $\Xo_{\mu\nu}$ is
antisymmetric under the exchange of $\mu \leftrightarrow \nu$, but
$\Xt_{\mu\nu}$ is symmetric under the same exchange.  Given the most general
effective Lagrangian for each of the spin possibilities, it is imperative that
one finds out the effective vertex factors from it. Using the effective vertex
factors one then proceeds to find out the angular distributions for each of the
spin possibilities, which are finally used to distinguish the different cases of
spin and parity.

Analyzing the effective Lagrangians given in Eqs.~\eqref{eq:Lagrangian-0},
\eqref{eq:Lagrangian-1} and \eqref{eq:Lagrangian-2} we can write down the
following vertex factors for each of the spin possibilities of $X$:
\begin{itemize}[leftmargin=4mm]
\item[$\bullet$]$J=0$:
\begin{equation}
\Vertex{\alpha\beta} = \Even{1}{0} \; g^{\alpha\beta} + \Even{2}{0} \;
P^{\alpha} P^{\beta} + i \; \Odd{1}{0} \; \epsilon^{\alpha\beta\rho\sigma}
q_{1\rho} q_{2\sigma},\label{eq:vertex-0}
\end{equation}
\item[$\bullet$]$J=1$:
\begin{equation}
\Vertex{\mu;\alpha\beta} = \Odd{1}{1} \; \left( g^{\alpha\mu} q_{1}^{\beta} +
g^{\beta\mu} q_{2}^{\alpha} \right) + i \; \Even{1}{1} \;
\epsilon^{\alpha\beta\mu\nu} \; Q_{\nu},\label{eq:vertex-1}
\end{equation}
\item[$\bullet$]$J=2$:
\begin{align}
\Vertex{\mu\nu;\alpha\beta} =& \Even{1}{2} \left(g^{\alpha\nu} \; g^{\beta\mu} +
g^{\alpha\mu} \; g^{\beta\nu}\right) \nn \\ %
& + \Even{2}{2} \Big( Q^{\mu} \left(Q^{\alpha} \; g^{\beta\nu} + Q^{\beta} \;
g^{\alpha\nu} \right) \nn \\ %
& \quad \qquad + Q^{\nu} \left(Q^{\alpha} \; g^{\beta\mu} + Q^{\beta} \;
g^{\alpha\mu}\right) \Big) \nn \\ %
& + \Even{3}{2} \left( Q^{\mu} \; Q^{\nu} \; g^{\alpha\beta} \right) -
\Even{4}{2} \left( Q^{\alpha} \; Q^{\beta} \; Q^{\mu} \; Q^{\nu} \right) \nn\\ %
& + 2 i \; \Odd{1}{2} \Big( g^{\beta\nu} \; \epsilon^{\alpha\mu\rho\sigma} -
g^{\alpha\nu} \; \epsilon^{\beta\mu\rho\sigma} \nn \\ %
& \quad \qquad \quad + g^{\beta\mu} \; \epsilon^{\alpha\nu\rho\sigma} -
g^{\alpha\mu} \;\epsilon^{\beta\nu\rho\sigma} \Big) q_{1\rho} q_{2\sigma}\nn\\ %
& + i \; \Odd{2}{2} \Big( Q^{\beta} \left( Q^{\nu} \;
\epsilon^{\alpha\mu\rho\sigma} + Q^{\mu} \; \epsilon^{\alpha\nu\rho\sigma}
\right) \nn\\ %
& \quad \qquad - Q^{\alpha} \left( Q^{\nu} \; \epsilon^{\beta\mu\rho\sigma} +
Q^{\mu}\;\epsilon^{\beta\nu\rho\sigma}\right) \Big) q_{1\rho} q_{2\sigma}\nn\\ %
& + i \; \Odd{3}{2} \Big( \epsilon^{\alpha\beta\nu\rho} P_{\rho} Q^{\mu} +
\epsilon^{\alpha\beta\mu\rho} P_{\rho} Q^{\nu} \Big) \nn \\ %
& + i \; \Odd{4}{2} \; \epsilon^{\alpha\beta\rho\sigma} \; Q^{\mu} \; Q^{\nu} \;
q_{1\rho} q_{2\sigma},\label{eq:vertex-2}
\end{align}
\end{itemize}
where $\Even{i}{J}, \Odd{i}{J}$ are the form factors which are associated with
terms that are even and odd under parity respectively, and are related to the
effective coupling constants $\even{i}{J}, \odd{i}{J}$ (as given in
\ref{sec:appendix-form-factors-coupling-constants}). It is important to notice
that the terms proportional to the Levi-Civita tensor
$\epsilon^{\alpha\beta\mu\nu}$ are the terms that are odd under parity in the
vertex factors for spin $0$ and $2$, but not in case of spin $1$. This can be
easily explained by the fact that in case of spin $1$ there are three
polarization vectors that come into picture, which in association with the
Levi-Civita tensor and the momentum $Q$ give rise to a triple product. This
triple product involves the three polarizations and is even under parity. Thus
the term with Levi-Civita tensor carries the even form factor in case of spin
$1$ instead of the usual odd form factor. Moreover, the imaginary $i$ remains
with the Levi-Civita tensor to keep the vertex factor even under time reversal.

It is noteworthy that in Ref.~\cite{Modak:2013sb}, the notation used is slightly
different from the ones used here. In order to go to the notation used in
Ref~\cite{Modak:2013sb}, the necessary substitutions are tabulated in
Table~\ref{relation}.
\begin{table}[hbtp]
\centering
\begin{tabular}{>{\centering\arraybackslash}p{2.2cm} >{\centering\arraybackslash}p{2.2cm}} \hline %
Spin 0 & Spin 2 \\ \hline %
$\Even{1}{0} \to \displaystyle \frac{i g M_Z}{\cos\theta_W} a$ & $\Even{1}{2}
\to A$ \\ %
$\Even{2}{0} \to \displaystyle \frac{i g M_Z}{\cos\theta_W} b$ & $\Even{2}{2}
\to B$ \\ %
$\Odd{1}{0} \to \displaystyle \frac{i g M_Z}{\cos\theta_W} c$ & $\Even{3}{2} \to
C$\\ %
& $\Even{4}{2} \to D$\\ %
& $\Odd{1}{2} \to E$\\ %
& $\Odd{2}{2} \to F$\\ %
& $\Odd{3}{2} \to G$\\ \hline %
\end{tabular}
\caption{The correspondence between the vertex factors of this paper with those
given in Ref.~\cite{Modak:2013sb}.} %
\label{relation}
\end{table}

The Lagrangians and vertex factors for the different spin possibilities have
also been considered in the literature before~\cite{Miller:2001bi, Choi:2002dq,
Choi:2002jk, Gao:2010qx, Keung:2008ve, Cheung:2011nv, Giudice:1998ck,
Hagiwara:2008jb, Artoisenet:2013puc}. Our approach differs from these works by
isolating the parity-even and odd terms explicitly, such that the uniangular
distributions can be profitably used for characterizing the spin and parity of
the particle. It is important to notice that there is no contribution from
$\odd{2}{1}$ and $\even{2}{1}$ to the vertex factors. In order to understand
this let us have a look at the terms $\Xo_{\mu\nu}$ and $\Xot_{\mu\nu}$ which
come in association with $\odd{2}{1}$ and $\even{2}{1}$ respectively. Under $\mu
\leftrightarrow \nu$ exchange both $\Xo_{\mu\nu}$ and $\Xot_{\mu\nu}$ pick up a
relative negative sign. However, the $Z^{\mu} Z^{\nu}$ part is symmetric under
the $\mu \leftrightarrow \nu$ exchange, as it should be for the case of two
identical $Z$ bosons. So considering Bose symmetry of the two daughter $Z$
bosons, it is clear that the terms $\odd{2}{1}$ and $\even{2}{1}$ cannot
contribute, as these are overall anti-symmetric under Bose symmetry.

Before we embark upon the journey to find out the differential decay rate or the
uni-angular distributions, it would be nice to have some physical understanding
of the process under consideration. Analysing the decay $X \to ZZ$ in terms of
the helicity amplitudes and partial waves offer valuable insight to the
understanding of the process.

\subsection{Helicity amplitudes in the transversity basis for the decay \texorpdfstring{$X \to ZZ$}{X->ZZ}} \label{subsec:transversity-amplitudes}

Helicity amplitudes in the transversity basis in which we are going to work have
a very special property: the helicity amplitudes are now sensitive to the parity
of the resonance $X$. Thus our helicity amplitudes have distinct and unambiguous
parity signatures. We shall denote those helicity amplitudes which have
parity-even form factors as $\Ampe{i}{J}$ and those with parity-odd form factors
as $\Ampo{i}{J}$, where the superscript $(J)$ denotes the spin of the parent
particle $X$. The formalism for the helicity amplitudes is discussed in the
~\ref{sec:helicity-partial-wave}.  Ignoring the lepton masses in the final
state, we get the following helicity amplitudes for the different spin
possibilities of $X$:
\begin{itemize}[leftmargin=4mm]
\item[$\bullet$]$J=0$:
\begin{subequations} \label{eq:Helicity-amplitudes-spin0}
\begin{align}
\Ampe{1}{0} &= \frac{1}{2} \left( M_X^2 - M_1^2 - M_2^2 \right) \, \Even{1}{0} +
M_X^2 Y^2 \, \Even{2}{0}, \\ %
\Ampe{2}{0} &= \sqrt{2} M_1 M_2 \, \Even{1}{0}, \\ %
\Ampo{1}{0} &= \sqrt{2} M_1 M_2 M_X Y \, \Odd{1}{0},
\end{align}
\end{subequations}
\item[$\bullet$]$J=1$:
\begin{subequations} \label{eq:Helicity-amplitudes-spin1}
\begin{align}
\Ampo{1}{1} &= \frac{\sqrt{2}}{3} D_1 \, \Odd{1}{1}, \\ %
\Ampe{1}{1} &= \frac{1}{3} D_2 \, \Even{1}{1},
\end{align}
\end{subequations}
\item[$\bullet$]$J=2$:
\begin{subequations} \label{eq:Helicity-amplitudes-spin2}
\begin{align}
\Ampe{1}{2} &= \frac{2 \sqrt{2}}{3 \sqrt{3} M_X^2} \bigg( \Even{1}{2} \left(
M_X^4 - M_{-}^4 \right) - \Even{2}{2} \left( 8 M_X^4 Y^2 \right) \nn \\ %
& \qquad \qquad \qquad + \Even{3}{2} \left( 4M_X^2 Y^2 \right) \left( M_{+}^2 -
M_X^2 \right) \nn\\ %
& \qquad \qquad \qquad - \Even{4}{2} \left( 8 M_X^4 Y^4 \right) \bigg), \\ %
\Ampe{2}{2} &= \frac{8 M_1 M_2}{3 \sqrt{3}} \left( \Even{1}{2} + 4 Y^2
\Even{3}{2} \right), \\ %
\Ampe{3}{2} &= \frac{4}{3 M_X M_{+}} \bigg( \Even{1}{2} \left( M_{-}^4 - M_X^2
M_{+}^2 \right) \nn\\ %
& \qquad \qquad \qquad + \Even{2}{2} \left( 4 M_{+}^2 M_X^2 Y^2 \right) \bigg),
\\ %
\Ampe{4}{2} &= \frac{8 M_1 M_2 \nu^2}{3 M_X M_{+}} \; \Even{1}{2},\\ %
\Ampo{1}{2} &= \frac{4 Y}{3 M_{+}} \bigg( \Odd{1}{2} \left( M_{-}^4 - M_X^2
M_{+}^2 \right) + \Odd{2}{2} \left( 4 M_{+}^2 M_X^2 Y^2 \right) \bigg), \\ %
\Ampo{2}{2} &= \frac{8 M_1 M_2 \mu^2 Y}{3 \sqrt{3} M_{+}} \; \Odd{1}{2} ,
\end{align}
\end{subequations}
\end{itemize}
where $M_X$ is the mass of the resonance $X$, $M_1$ is the invariant mass of the
$\ell_1^{\pm}$ lepton pair, $M_2$ is the invariant mass of the $\ell_2^{\pm}$
lepton pair (for off-shell contributions $M_1^2$, $M_2^2$ are not equal to
$M_Z^2$), $Y$ is the magnitude of the three-momentum with which the two $Z$
bosons fly away back-to-back in the rest frame of $X$:
\begin{equation}\label{eq:Y}
Y = \frac{\sqrt{\lambda(M_X^2, M_1^2, M_2^2)}}{2 M_X},
\end{equation}
with
\begin{equation}\label{eq:Kallen-function}
\lambda(x,y,z) = x^2 + y^2 + z^2 - 2 (x y+ yz + zx),
\end{equation}
being the K\"{a}ll\'{e}n function, $D_1$ and $D_2$ being given by
\begin{align}
D_1^2 &= 2 M_X^6 M_{+}^2 - \frac{1}{2} M_X^4 \left(5 M_{+}^4 + M_{-}^4 \right)
\nn\\ %
& \qquad + 6 M_X^2 \left( M_{+}^6 - M_{+}^2 M_{-}^4 \right) + \frac{3}{2}
M_{+}^4 M_{-}^4 - \frac{1}{2} M_{-}^8,\\ %
D_2^2 &=16 M_X^6 M_{+}^2+56 M_X^4 M_{+}^4-86 M_X^2 M_{+}^6 \nn\\ %
& \qquad +M_{-}^4 \left(-85 M_X^4+96 M_X^2 M_{+}^2-35 M_{+}^4\right)+38 M_{-}^8,
\end{align}
with $M_{+}$, $M_{-}$, $\mu^2$ and $\nu^2$ being given by
\begin{align}
M_{+}^2 &= M_1^2 + M_2^2, \\ %
M_{-}^2 &= M_1^2 - M_2^2, \\ %
\mu^4 &= 4 M_X^2 M_{+}^2 + 3 M_{-}^4, \\ %
\nu^4 &= 2 M_X^2 M_{+}^2 + M_{-}^4.
\end{align}
It is noteworthy that considering the vertex factors with form factors
$\Even{1}{2}$, $\Even{2}{2}$, $\Even{3}{2}$, $\Even{4}{2}$, $\Odd{1}{2}$ and
$\Odd{2}{2}$ suffice to provide the most general angular distribution for the
spin-$2$ case. Including the form factors $\Odd{3}{2}$, and $\Odd{4}{2}$ result
in only a redefinition of the form factors $\Odd{1}{2}$ and $\Odd{2}{2}$ as
discussed in \ref{sec:appendix-redundancy-in-spin-2-vertex}.

It is again noteworthy that the amplitudes used here are notationally different
from the amplitudes used in Ref.~\cite{Modak:2013sb} for better clarity of their
physical content. By following the substitutions as given in
Table~\ref{tab:amplitudes-old-new} one can go back to the notations used in
Ref.~\cite{Modak:2013sb}. The notations $M_{\pm}$, $\mu$ and $\nu$ have been
introduced to make the power counting of mass dimensions easy.

\begin{table}[hbtp]
\centering
\begin{tabular}{>{\centering\arraybackslash}p{2.2cm}
>{\centering\arraybackslash}p{2.2cm}} \hline %
Spin 0 & Spin 2 \\ \hline %
$\Ampe{1}{0} \to A_L$ & $\Ampo{1}{2} \to A_L$ \\ %
$\Ampe{2}{0} \to A_{\parallel}$ & $\Ampo{2}{2}\to A_M$ \\ %
$\Ampo{1}{0} \to A_{\perp}$ & $\Ampe{1}{2} \to A_1$ \\ %
 & $\Ampe{2}{2} \to A_2$ \\ %
 & $\Ampe{3}{2} \to A_3$ \\ %
 & $\Ampe{4}{2} \to A_4$ \\ %
 & $Y \to X$ \\ %
 & $M_{+} \to \mathsf{u}_1$ \\ %
 & $M_{-} \to \mathsf{u}_2$ \\ %
 & $\mu^2 \to \mathsf{v}$ \\ %
 & $\nu^2 \to \mathsf{w}$ \\ \hline
\end{tabular}
\caption{The correspondence between the helicity amplitudes of this paper and
those given in Ref.~\cite{Modak:2013sb}.} %
\label{tab:amplitudes-old-new}
\end{table}

\subsection{Uni-angular distributions for the decay \texorpdfstring{$X \to Z_1 Z_2 \to (\ell_1^+ \ell_1^-)(\ell_2^+\ell_2^-)$}{X -> ZZ -> l1+ l1- l2+ l2-}}
\label{subsec:uniangular}

Using the helicity amplitudes defined in
Eqs.~\eqref{eq:Helicity-amplitudes-spin0}, \eqref{eq:Helicity-amplitudes-spin1}
and \eqref{eq:Helicity-amplitudes-spin2}, the uni-angular distributions w.r.t
$\cos\theta_1$, $\cos\theta_2$ and $\phi$ can be written in a unified notation
for all the allowed spin possibilities of $X$ as follows:
\begin{align}
\frac{1}{\gammafJ} \frac{d^3\gammaJ}{dq_1^2 \, dq_2^2 \, d\cos\theta_1} &=
\frac{1}{2} + \Tone{J} \; \cos\theta_1 + \Ttwo{J} \; P_2(\cos\theta_1),
\label{eq:cost1} \\ %
\frac{1}{\gammafJ} \frac{d^3\gammaJ}{dq_1^2 \, dq_2^2 \, d\cos\theta_2} &=
\frac{1}{2} + \Tonep{J} \cos\theta_2 + \Ttwop{J}
P_2(\cos\theta_2),\label{eq:cost2}\\ %
\frac{2\pi}{\gammafJ} \frac{d^3\gammaJ}{dq_1^2 \, dq_2^2 \, d\phi} &= 1 +
\Uone{J} \; \cos\phi + \Utwo{J} \; \cos2\phi \nn\\ %
& \quad + \Vone{J} \; \sin\phi + \Vtwo{J} \; \sin2\phi, \label{eq:phi}
\end{align}
where $\Ti{J}$, $\Tip{J}$, $\Ui{J}$, and $\Vi{J}$ are the coefficients of
$P_i(\cos\theta_1)$, $P_i(\cos\theta_2)$, $\cos(i\phi)$, and $\sin(i\phi)$
respectively with $P_i(x)$ being the $i$'th Legendre polynomial in $x$. These
coefficients can easily be obtained from the uni-angular distributions by means
of the following asymmetries:
\begin{align}
  \Tone{J} &= \left( -\int_{-1}^{0} + \int_{0}^{+1} \right)
  d\cos\theta_1 \; \left(
  \frac{1}{\gammafJ}\frac{d^3\gammaJ}{dq_1^2 \; dq_2^2 \;
    d\cos\theta_1} \right),\label{eq:T1Jasym}\\
  \Ttwo{J} &= \frac{4}{3} \left( \int_{-1}^{-\frac{1}{2}} -
  \int_{-\frac{1}{2}}^{0} - \int_{0}^{+\frac{1}{2}} +
  \int_{+\frac{1}{2}}^{+1} \right) d\cos\theta_1 \nn\\
&\quad\quad \quad\quad \times\; \left(
  \frac{1}{\gammafJ}\frac{d^3\gammaJ}{dq_1^2 \; dq_2^2 \;
    d\cos\theta_1} \right),\label{eq:T2Jasym}\\
  \Tonep{J} &= \left( -\int_{-1}^{0} + \int_{0}^{+1} \right)
  d\cos\theta_2 \; \left(
  \frac{1}{\gammafJ}\frac{d^3\gammaJ}{dq_1^2 \; dq_2^2 \;
    d\cos\theta_2} \right),\label{eq:T1pJasym}\\
  \Ttwop{J} &= \frac{4}{3} \left( \int_{-1}^{-\frac{1}{2}} -
  \int_{-\frac{1}{2}}^{0} - \int_{0}^{+\frac{1}{2}} +
  \int_{+\frac{1}{2}}^{+1} \right) d\cos\theta_2 \nn\\
&\quad\quad \quad\quad \times\; \left(
  \frac{1}{\gammafJ}\frac{d^3\gammaJ}{dq_1^2 \; dq_2^2 \;
    d\cos\theta_2} \right),\label{eq:T2pJasym}\\
  \Uone{J} &= \frac{1}{4} \left( - \int_{-\pi}^{-\frac{\pi}{2}} +
  \int_{-\frac{\pi}{2}}^{+\frac{\pi}{2}} -
  \int_{+\frac{\pi}{2}}^{+\pi} \right) d\phi \; \left(
  \frac{2\pi}{\gammafJ}\frac{d^3\gammaJ}{dq_1^2 \; dq_2^2 \;
    d\phi} \right),\label{eq:U1Jasym}\\
  \Utwo{J} &= \frac{1}{4} \left( \int_{-\pi}^{-\frac{3\pi}{4}} -
  \int_{-\frac{3\pi}{4}}^{-\frac{\pi}{4}} +
  \int_{-\frac{\pi}{4}}^{+\frac{\pi}{4}} -
  \int_{+\frac{\pi}{4}}^{+\frac{3\pi}{4}} +
  \int_{+\frac{3\pi}{4}}^{+\pi} \right) d\phi \nn\\
&\quad\quad \quad \times\; \left(
  \frac{2\pi}{\gammafJ}\frac{d^3\gammaJ}{dq_1^2 \; dq_2^2 \;
    d\phi} \right),\label{eq:U2Jasym}\\
  \Vone{J} &= \frac{1}{4} \left( - \int_{-\pi}^{0} + \int_{0}^{+\pi}
  \right) d\phi \; \left(
  \frac{2\pi}{\gammafJ}\frac{d^3\gammaJ}{dq_1^2 \; dq_2^2 \;
    d\phi} \right),\label{eq:V1Jasym}\\
  \Vtwo{J} &= \frac{1}{4} \left( \int_{-\pi}^{-\frac{\pi}{2}} -
  \int_{-\frac{\pi}{2}}^{0} + \int_{0}^{+\frac{\pi}{2}} -
  \int_{+\frac{\pi}{2}}^{+\pi} \right) d\phi \nn\\
&\quad\quad  \times \; \left(
  \frac{2\pi}{\gammafJ}\frac{d^3\gammaJ}{dq_1^2 \; dq_2^2 \;
    d\phi} \right).\label{eq:V2Jasym}
\end{align}
The differential decay width $\gammafJ$ is defined as
\begin{equation}
  \gammafJ = \frac{d^2 \gammaJ}{dq_1^2 \, dq_2^2} =
  \NJ \left( \sum_i \modulus{\Ampe{i}{J}}^2 + \sum_j
  \modulus{\Ampo{j}{J}}^2 \right)
\end{equation}
where the factor $\NJ$ is given by
\begin{align} \label{eq:NJ}
  \NJ = \SJ \frac{9}{2^{10}} \frac{1}{\pi^4}
  \frac{\Br_{\ell\ell}^2}{M_X^2} \frac{\Gamma_Z^2}{M_Z^2} Y
  \frac{1}{ \left( q_1^2 - M_Z^2 \right)^2 + M_Z^2 \Gamma_Z^2 } \nn \\
    \times\frac{1}{ \left( q_2^2 - M_Z^2 \right)^2 + M_Z^2
      \Gamma_Z^2 },
\end{align}
with $\SJ$ being the factor that comes from summing over initial spin
states, i.e.\ $\SJ = 1, \frac{1}{3}, \frac{1}{5}$ for scalar, vector
and tensor resonances respectively, $\Gamma_Z$ being the total decay
width of the $Z$ boson, $\Br_{\ell\ell}$ being the branching ratio for
the decay of $Z$ to a pair of charged leptons: $Z \to \ell^+
\ell^-$. The coefficients $\Ti{J}$, $\Tip{J}$, $\Ui{J}$, $\Vi{J}$
introduced in Eqs.~\eqref{eq:cost1}, \eqref{eq:cost2}, \eqref{eq:phi}
are expressed in terms of \textit{helicity fractions} $\Fe{i}{J}$ and
$\Fo{i}{J}$ which are defined as
\begin{equation}\label{eq:Helicity-fractions}
\left.
\begin{aligned}
\Fe{i}{J} &= \frac{\Ampe{i}{J}}{\displaystyle \sum_i
  \modulus{\Ampe{i}{J}}^2 + \sum_j \modulus{\Ampo{j}{J}}^2},\\
 \Fo{i}{J} &= \frac{\Ampo{i}{J}}{\displaystyle \sum_i
   \modulus{\Ampe{i}{J}}^2 + \sum_j \modulus{\Ampo{j}{J}}^2},
\end{aligned} \right\}
\end{equation}
such that
\begin{equation}
  \sum_i \modulus{\Fe{i}{J}}^2 + \sum_j \modulus{\Fo{j}{J}}^2 = 1.
\end{equation}
The expressions for the coefficients $\Ti{J}$, $\Tip{J}$, $\Ui{J}$, $\Vi{J}$ are
given in \ref{sec:apendix-TTUV}.  Looking at the uni-angular distributions it is
easy to find that for any spin case, the coefficients $\Tone{J}$, $\Tonep{J}$,
$\Vone{J}$ and $\Vtwo{J}$ are the interference terms between the parity-even and
parity-odd terms. So if the resonance $X$ were to be a parity eigenstate, these
interference terms must vanish irrespective of the spin of $X$. Therefore, the
conditions for $X$ to be a parity eigenstate are
\begin{equation}
  \label{eq:parity-eigenstate-cond}
  \Tone{J} = \Tonep{J} = \Vone{J} = \Vtwo{J} = 0.
\end{equation}
Now it is again easy to observe that for spin-$0$ case the coefficients
$\Ttwo{0}$ and $\Ttwop{0}$ are the same. However this is not true for spin-$1$
and spin-$2$ cases, in general. In order to further analyse this situation we
define the difference $\Delt{J}$ between $\Ttwo{J}$ and $\Ttwop{J}$:
\begin{equation}
  \label{eq:delta}
  \Delt{J} = \Ttwo{J} - \Ttwop{J}.
\end{equation}
For the different spin possibilities we have
\begin{align}
  \Delt{0} &= 0,\\
  \Delt{1} &= 6 M_X^2 M_{-}^2 Y^2 \Bigg(\frac{\modulus{\Fo{1}{1}}^2
    \left(M_X^2+M_{+}^2\right)}{D_1^2} \nn\\
& \qquad \qquad \qquad \quad + \frac{2 \modulus{\Fe{1}{1}}^2
    \left(5 M_X^2+M_{+}^2\right)}{D_2^2}\Bigg),\\
  \Delt{2} &= \frac{3M_{-}^2}{4 M_{+}^2 \mu^4 \nu^4} \bigg( \mu^4
  \nu^4 \left( \modulus{\Fe{3}{2}}^2 + \modulus{\Fo{1}{2}}^2 \right)
  \nn \\
  & \qquad \qquad - M_{-}^4 \left( \mu^4 \modulus{\Fe{4}{2}}^2 + 3 \,
  \nu^4 \modulus{\Fo{2}{2}}^2 \right) \bigg) \nonumber \\
  &+\frac{3M_1 M_2 M_{-}^2}{M_{+}^2 \mu^2 \nu^2} \bigg( \mu^2 \,
  \Re{\left(\Fe{3}{2} \Fest{4}{2}\right)} + \sqrt{3} \, \nu^2 \,
  \Re{\left(\Fo{1}{2} \Fost{2}{2}\right)}\bigg).
\end{align}
It is clear that we can never have $\Delt{1} = 0$ for all values of $M_1$ and
$M_2$, since this requires $\Fe{1}{1} = \Fo{1}{1} =0$, which is an unphysical
condition to satisfy as it totally obliterates the vertex factor itself.
Requiring that $\Delt{2} = 0$ for all values of $M_1$ and $M_2$ implies that
$\Fe{3}{2} = \Fe{4}{2} = \Fo{1}{2} = \Fo{2}{2} = 0$, which in turn implies that
$\Even{1}{2} = \Even{2}{2} = \Odd{1}{2} = \Odd{2}{2} =0$. Thus it leaves out two
form factors in the vertex factor, namely $\Even{3}{2}$ and $\Even{4}{2}$. This
case constitutes a very special case of the spin $2$ scenario and all its
uni-angular distributions are completely indistinguishable from the
corresponding angular distributions for a parity even spin-$0$ resonance. A
closer look at the helicity amplitudes for these two cases, $J^P = 0^+$ and
special $2^+$, reveals that the special $2^+$ case has extra $Y^2$ dependence
when compared with $0^+$ case:
\begin{align}
  \Ampe{1}{0} &= \frac{1}{2} \left( M_X^2 - M_1^2 - M_2^2 \right) \,
  \Even{1}{0} + M_X^2\, Y^2 \, \Even{2}{0},\\
  \Ampe{2}{0} &= \sqrt{2} \, M_1 \,M_2 \, \Even{1}{0} ,\\
  \Ampe{1}{2} &= \left(-\frac{16 \sqrt{2}}{3 \sqrt{3}} \; Y^2 \right)
  \; \bigg( \frac{1}{2} \left( M_X^2 - M_1^2 - M_2^2 \right) \,
  \Even{3}{2} \nn\\
& \qquad \qquad \qquad \qquad \qquad \qquad + M_X^2 Y^2 \, \Even{4}{2}
  \bigg), \\
  \Ampe{2}{2} &= \left(\frac{16\sqrt{2}}{3 \sqrt{3}} \; Y^2 \right) \;
  \sqrt{2} \, M_1 \, M_2 \, \Even{3}{2}.
\end{align}
\begin{figure*}[hbtp]
  \centering \includegraphics[scale=0.9]{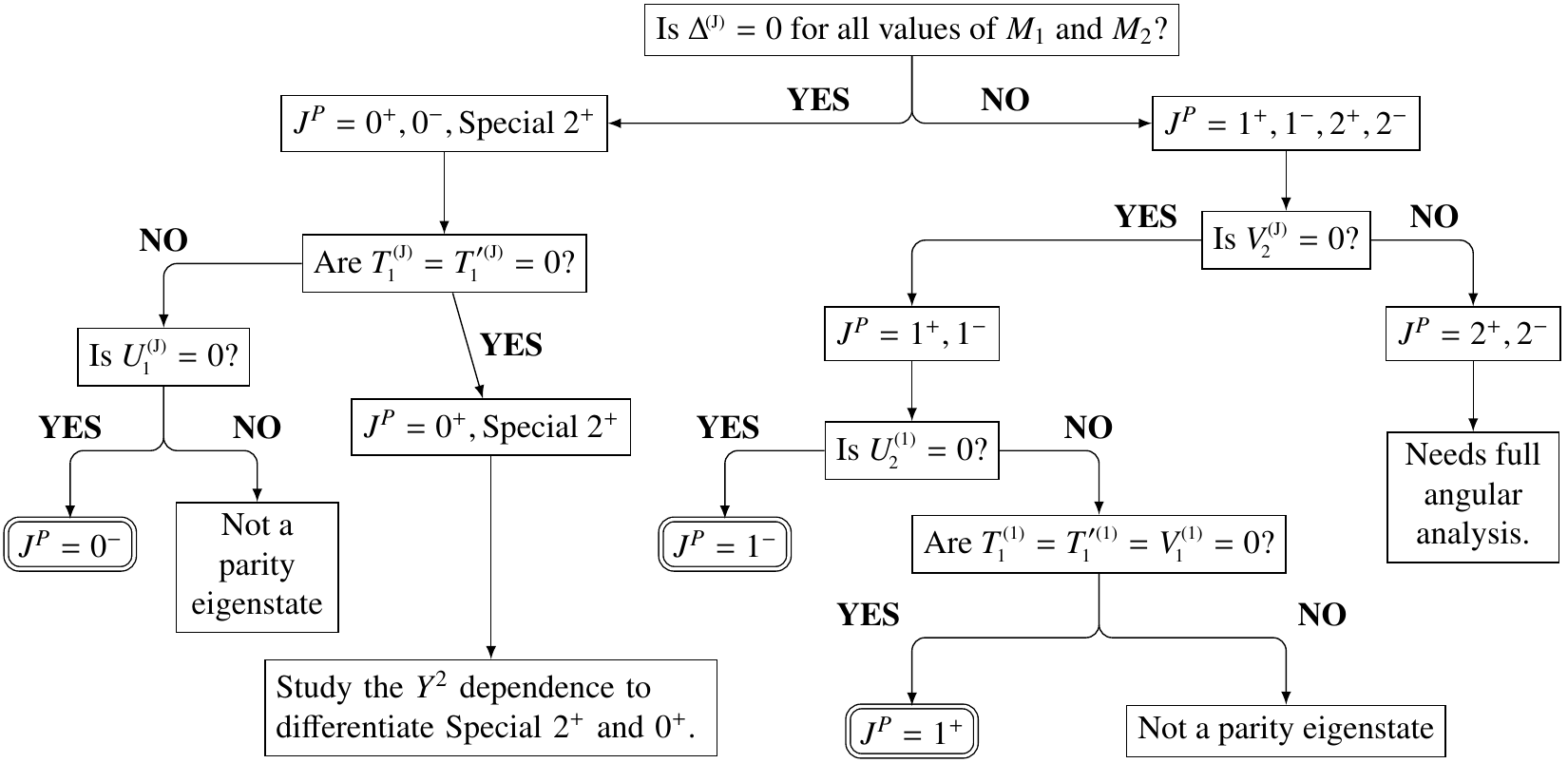} \caption{Flowchart
  to determine the spin and parity of a resonance $X$ decaying as $X \to Z_1 Z_2
  \to (\ell_1^- \ell_1^+)(\ell_2^- \ell_2^+)$.} \label{fig:flowchart}
\end{figure*}
Looking at the spin-$1$ case we find that the coefficient $\Utwo{1}$ is
sensitive to $\Even{1}{1}$. And hence this term can be used to identify the
parity of the spin-$1$ resonance.  We can thus find out a step-by-step
methodology that can be followed to uniquely predict the spin and parity of a
resonance decaying to four final charged leptons via two $Z$ bosons. This is
given in Fig.~\ref{fig:flowchart}.

\section{Numerical Study}\label{sec:numerical}

In this section we will show how the uniangular distributions given in
Eqs.~\eqref{eq:cost1}, \eqref{eq:cost2} and \eqref{eq:phi} can be used to find
out the values of the angular observables defined in Eqs.~\eqref{eq:T1Jasym},
\eqref{eq:T2Jasym}, \eqref{eq:T1pJasym}, \eqref{eq:T2pJasym},
\eqref{eq:U1Jasym}, \eqref{eq:U2Jasym}, \eqref{eq:V1Jasym} and
\eqref{eq:V2Jasym}. We shall elucidate the methodology by concentrating on some
heavy spin $1$ resonances and study the observables for them. We start by
investigating how the mass and decay width of such resonances affect their
production cross-section in the future LHC runs. We then benchmark the angular
observables for spin-$1^{+}$ and spin-$1^{-}$ resonances for two different
Center-of-Momentum (CM) energies: $14~\tev$ and $33~\tev$ with 3000 fb$^{-1}$
luminosity. This analysis can be easily extended to consider a heavy spin-2
resonance.
  
Let us consider a heavy spin-$1$ resonance $X$ of mass $M_X$ and decay width
$\Gamma_X$, decaying into four charged leptons via two $Z$ bosons. We shall
assume that the resonance $X$ is produced via annihilation of quark ($q$) and
antiquark ($\bar q$) pairs. The production process is characterized by the
effective Lagrangian,
\begin{equation}\label{eq:prod}
\mathscr{L}_{eff} = \sum_q \left( \tilde{c}_q \; \bar{q} \gamma^{\mu} q\;
X_{\mu} +  c_q \; \bar{q} \gamma^{\mu} \gamma^5 q\; X_{\mu} \right),
\end{equation}
where $q= u, d, c, s, b $ quarks and $\tilde{c}_q$, $c_q$ are the coupling
strengths of $X$ to vector, axial vector currents respectively; i.e.\ for a spin
$1^{+}$ resonance $\tilde{c}_q=0$ and for a spin $1^{-}$ resonance $c_q=0$. Just
for simplicity of the analysis we have further assumed that all the quarks
couple to the resonance $X$ with the same strength. The production cross section
for the resonance $X$ can be easily obtained from Eq.~\eqref{eq:prod} by
considering appropriate parton distribution functions in the process $p p \to
X$. The production cross section does depend on the mass of the resonance $X$:
larger the mass, lower is the production cross section at a given CM energy.

The decay process $X \to Z Z $ is characterized by the Lagrangian
Eq.~\eqref{eq:Lagrangian-1}. The partial decay width for $X \to Z Z$ for
spin-$1$ resonance is given by
\begin{equation}
\Gamma_{ZZ} = O_1^2 \frac{M_X^3}{32 M_Z^2 \pi}\left({1- \frac{4
M_Z^2}{M_X^2}}\right)^{\frac{3}{2}}+ E_1^2 \frac{M_X^3}{32 M_Z^2 \pi}\left({1-
\frac{4 M_Z^2}{M_X^2}}\right)^{\frac{5}{2}}.
\end{equation}
For a spin-$1^{+}$ resonance $O_1=0$ and for a spin-$1^{-}$ resonance $E_1=0$.
It must be noted that we have dropped the superscript ``(1)'' from both $E_1$
and $O_1$ throughout this section.
 
\begin{figure}[htb!]
  \centering \includegraphics[scale=0.8]{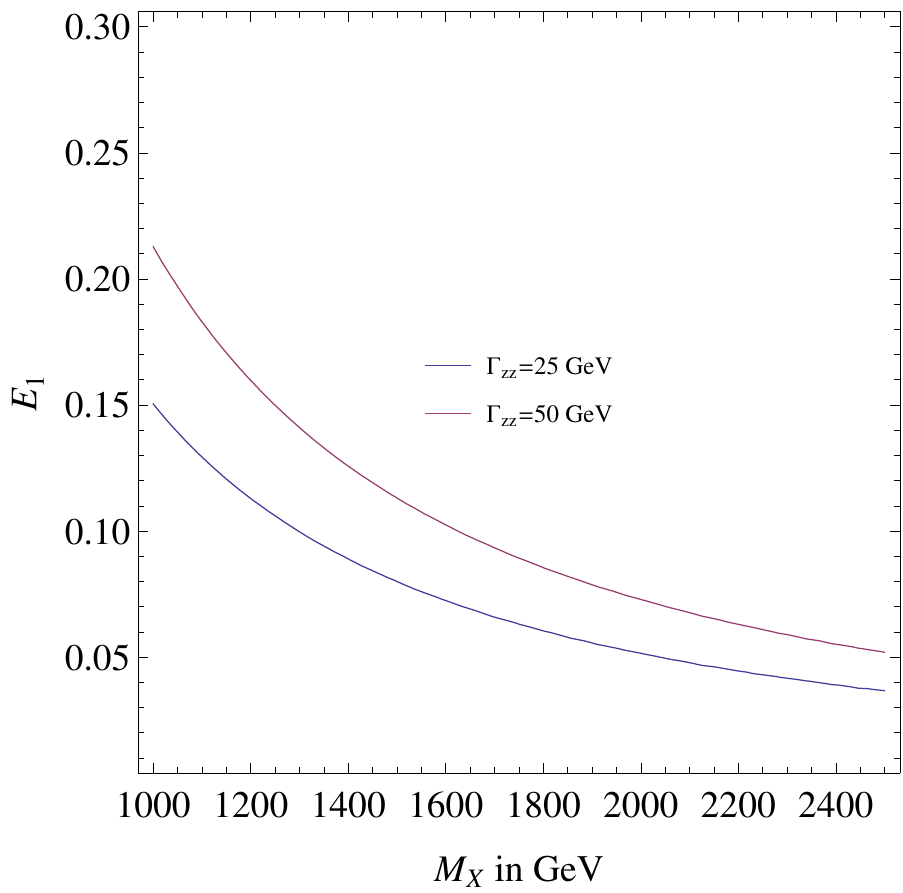} \caption{Mass ($M_X$) vs.
  $E_1$ plot of a Spin $1^+$ resonance for different $\Gamma_{ZZ}$. The blue
  curve for $\Gamma_{ZZ}=25$ and the purple curve for $\Gamma_{ZZ}=50$. }
  \label{fig:mxvsE01}
\end{figure}

The partial decay width for $X \to q \bar{q}$ for spin-$1$ resonance is given by
\begin{equation}
\Gamma_{q\bar{q}} = c_q^2 \frac{M_X}{4 \pi}\left({1 - \frac{4
m_q^2}{M_X^2}}\right)^{\frac{3}{2}}+ \tilde{c}_q^2 \frac{(M_X^2+2 m_q^2)}{4 M_X
\pi}\left({1- \frac{4 m_q^2}{M_X^2}}\right)^{\frac{1}{2}},
\end{equation}
where $m_q$ is the mass of the quark $q$ (or of antiquark $\bar q$), and $c_q=0$
for spin-$1^{-}$ resonance and $\tilde{c}_q=0$ for spin-$1^{+}$ resonance.  Let
us further assume that $X$ decays to all quark-antiquark pairs and to a pair of
$Z$ bosons only, i.e.\ the total decay width is given by
\begin{equation}
\Gamma_X= \Gamma_{ZZ}+\sum_q \Gamma_{q\bar{q}}\label{eq:decayofX}.
\end{equation}
One can relax these simple assumptions and do a detailed analysis where other
decay channels also exist. This will lead to modifications to the
Eq.~\eqref{eq:decayofX}. In Fig.~\ref{fig:mxvsE01} we show how the partial decay
width $\Gamma_{ZZ}$ varies with the mass $M_X$ of the spin-$1^+$ resonance, with
$M_Z \ll M_X$.
The spin-$1^-$ resonances also exhibit a similar plot for $M_Z \ll M_X$.

One possibility for a heavy spin-1 resonance is a heavy $Z'$ boson. The current
limit on the mass of a heavy $Z'$ resonance is
$1.7~\tev$~\cite{Khachatryan:2015sja}.  The current limit of $\tilde{c}_q$ and
$c_q$ for a particular mass $M_X$ of the resonance $X$, can be extracted out
from the $\sigma \times Br \times \mathcal{A}$ vs. resonance mass ($M_X$) plot
of Ref.~\cite{Khachatryan:2015sja}, where $\sigma$ is the cross-section for the
process $pp \to X$, $Br$ is the branching fraction of the decay $X \to q
\bar{q}$ and $\mathcal{A}$ is the acceptance.  Since the analysis of
Ref.~\cite{Khachatryan:2015sja} deals with search for a heavy resonance $Z'$
decaying to di-jet, which is a isotropic decay (two body final state), the
acceptance $\mathcal{A}$ is approximately $0.6$ and is independent of the mass
of $Z'$.

\begin{figure}[htb!]
  \centering \includegraphics[scale=0.5]{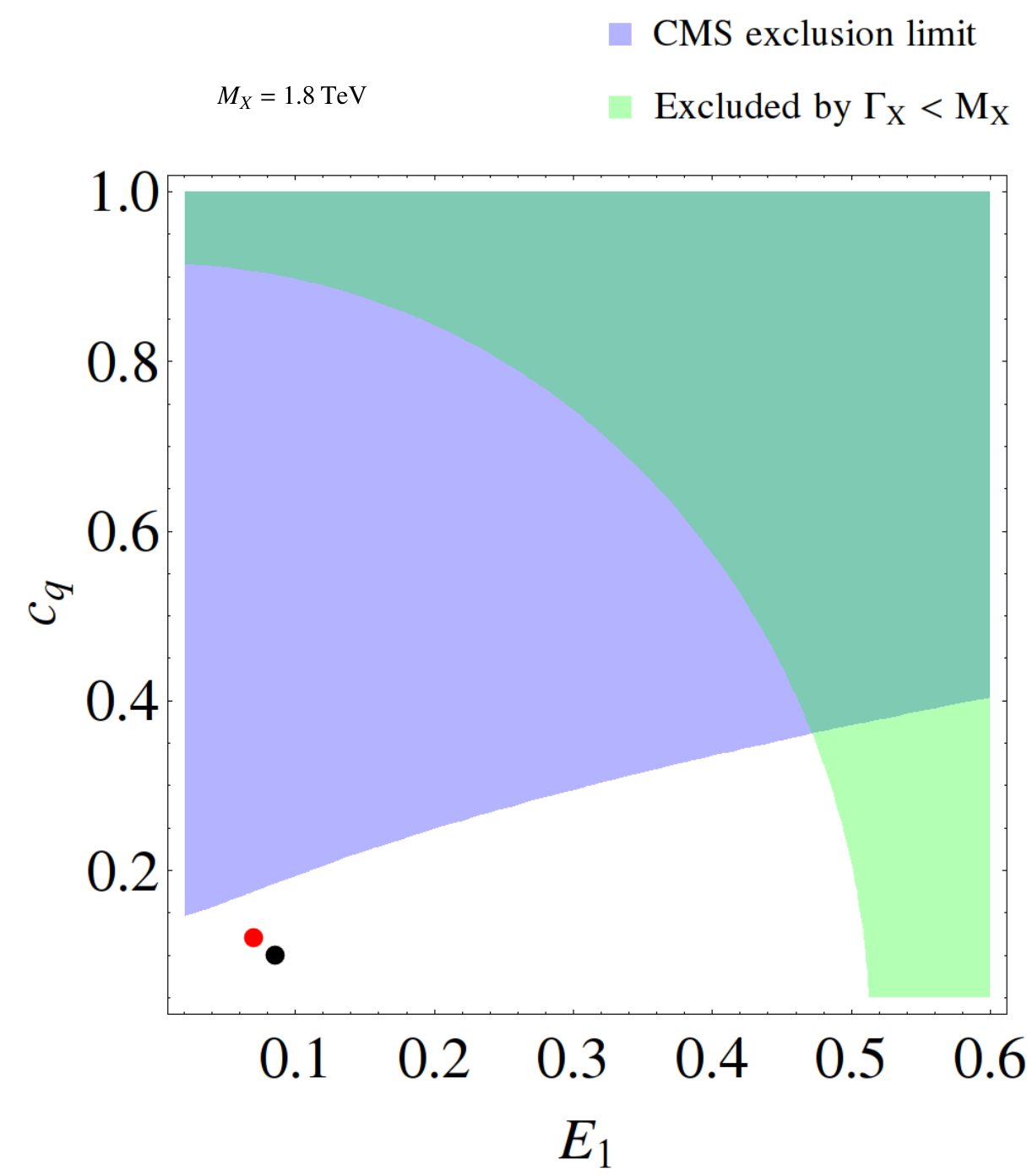} \caption{The allowed region
  for the couplings $c_q$ and $E_1$ for a spin-$1^+$ resonance of mass
  $M_X=1.8~\tev$. The green and the blue regions are excluded by $\Gamma_X <
  M_X$ limit and CMS limit from Ref.~\cite{Khachatryan:2015sja} respectively.
  The red ($E_1=7.00\times10^{-2} , c_q=0.12$) and the black
  ($E_1=8.56\times10^{-2} , c_q=0.10$) dots are the two benchmark points for our
  analysis.} \label{fig:cqe11p8}
\end{figure}

\begin{figure}[htb!]
  \centering \includegraphics[scale=0.5]{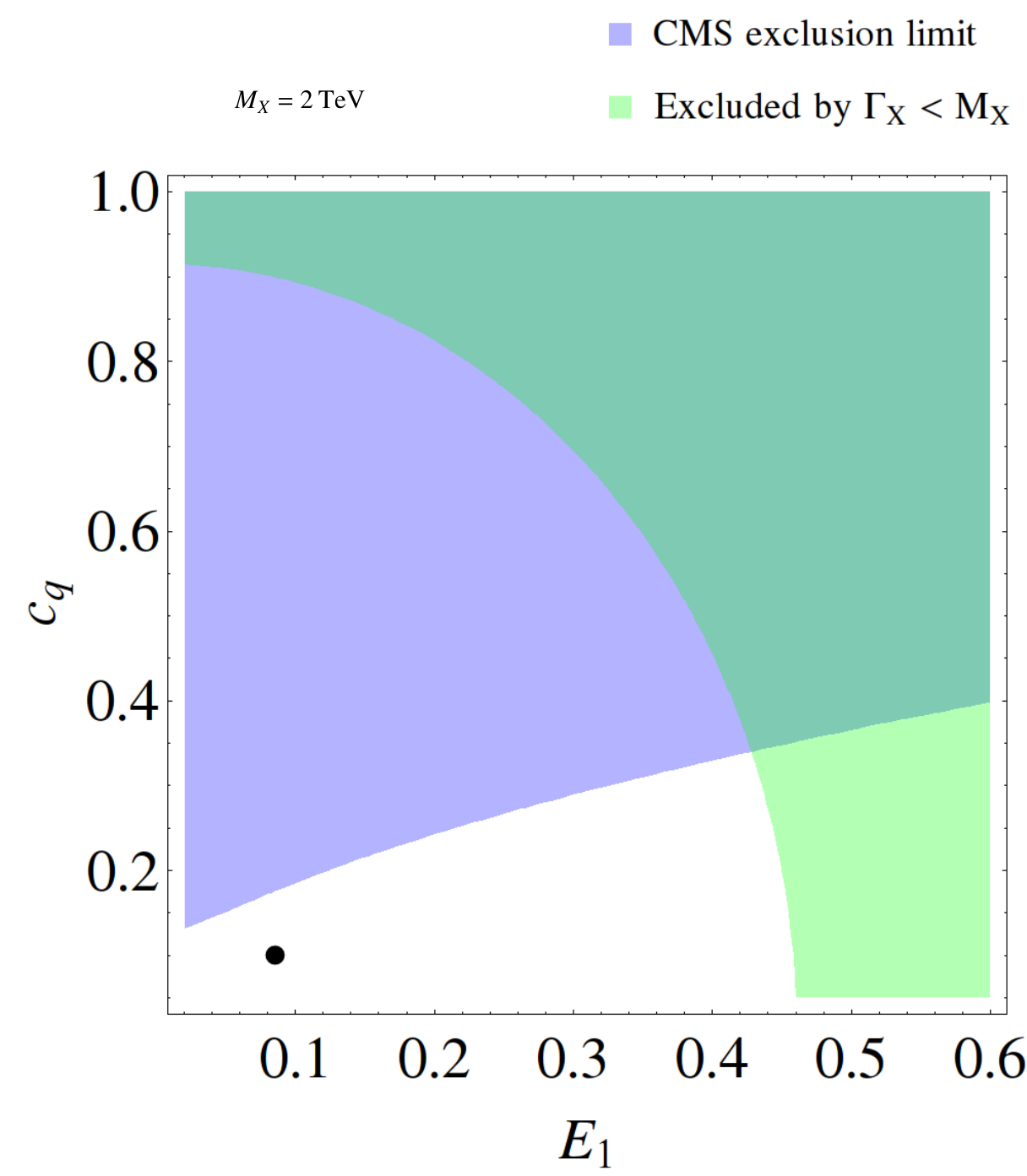} \caption{The allowed region
  for the couplings $c_q$ and $E_1$ for a spin-$1^+$ resonance of mass
  $M_X=2~\tev$. The green and the blue regions are excluded by $\Gamma_X < M_X$
  limit and CMS limit respectively. The black dot ($E_1=8.56\times10^{-2}
  ,~c_q=0.10$) is the benchmark point for our analysis. } \label{fig:cqe12}
\end{figure}

Following the analysis of Ref.~\cite{Khachatryan:2015sja} we find the allowed
region for the couplings $c_q$ and $E_1$ for two different masses, $M_X
=1.8~\tev$ and $2~\tev$, shown in Figs.~\ref{fig:cqe11p8} and \ref{fig:cqe12}
respectively.
From the allowed regions shown in Figs.~\ref{fig:cqe11p8} and \ref{fig:cqe12},
we choose three benchmark scenarios for masses $M_X= 1.8~\tev$ and $M_X=
2.0~\tev$ for our numerical study. The benchmark values of $c_q$ and $E_1$
corresponding to both the masses are tabulated in Table~\ref{param}.
\begin{table}[hbtp!]
\centering
\begin{tabular}{|c|c|c|c|}
\hline
Mass                           &Coupling~ $c_q$    & Coupling ~$E_1$  
        & $\Gamma_X$ in $\gev$\\
\hline
$1.8~\tev$                        & $0.12$            &  
$7.00\times10^{-2}$     &   $64.40$          \\
$1.8~\tev$                        & $0.10$            &  
$8.56\times10^{-2}$     &   $71.52$          \\
$2.0~\tev$                        & $0.10$            &  
$8.56\times10^{-2}$     &   $92.84$             \\
\hline
\end{tabular}
\caption{The benchmark values of the couplings $c_q$ and $E_1$, for a spin-$1^+$
resonance of masses $1.8~\tev$ and $2~\tev$ respectively for our analysis. The
values of corresponding decay widths $\Gamma_X$ are also tabulated in the last
column for both the masses.} \label{param}
\end{table}
Once the values of $M_X$, $c_q$(or $\tilde{c_q}$) and $E_1$(or $O_1$) for a
spin-$1^+$ (or spin-$1^-$) resonance are chosen, the total decay width
$\Gamma_X$ as well as the cross-section for the process $p p \to X \to Z Z \to
e^+ e^- \mu^+ \mu^-$ get fixed. The reader should note that, the process under
consideration is within the narrow width approximation where $\Gamma_X \ll M_X$.

For event generations we have used MADEVENT5~\cite{Alwall:2011uj} event
generator interfaced with PYTHIA6.4~\cite{Sjostrand:2003wg} and Delphes
3~\cite{deFavereau:2013fsa}.  The events are generated by $pp$ collisions via $q
\bar{q} \to X,$ for the CM energies $\sqrt{s}=$ $14~\tev$ and $33~\tev$, using
the parton distribution functions CTEQ6L1~\cite{Pumplin:2002vw}. Triggers as
well as electron and muon identification cuts are set following the analysis
presented in Refs.~\cite{atlash2zz,Aad:2009wy}.  We have only selected events
with final states $e^+ e^- \mu^+ \mu^-$, as our analysis is applicable only to
four non-identical final state leptons.  We have kept the trigger values same as
$14~\tev$, for the $33~\tev$ LHC analysis. However, it should be noted that in
the future $33~\tev$ LHC run, the trigger values may change, which could further
improve the statistics. The electron (muon) must satisfy $E_T>$ 7 $\gev$ ($p_T
>$ 6 $\gev$) and the pseudo-rapidity cut for electron (muon) is $\modulus{\eta}<
2.47$ ($\modulus{\eta} < 2.7$). The leptons are required to be separated from
each other by $\Delta R>0.1$ if they are of the same flavour and $\Delta R >0.2$
otherwise. The invariant mass cuts that are applied in our analysis are
$60~\gev<m_{ee}<120~\gev$, $60~\gev<m_{\mu \mu}<120~\gev$ and $1000~\gev<
m_{4\ell}$.

The effects of mass $M_X$ and width $\Gamma_X$ on $\sigma\times Br$ are shown in
Table~\ref{cut_table} for a spin-$1^{+}$ resonance.  The statistics lowers down
as the resonance gets heavier. However, the statistics improves for a resonance
with the same mass but narrower decay width.  This dependence is easily
discernible in Table~\ref{cut_table} for a spin-$1^+$ resonance. This mass and
width dependence on the cross-sections for the $p p \to X \to Z Z \to e^+ e^-
\mu^+ \mu^-$ process also shows the same behavior for a spin-$1^{-}$ resonance
in the limit $m_q \ll M_X$ and $M_Z \ll M_X$.
\begin{table*}[hbtp!]
 \centering
\begin{tabular}{|c|c|c|c|}
\hline
Cuts  in $\gev$                & $M_X=1.8~\tev$,               &
$M_X=1.8~\tev$,            & $M_X=2~\tev$,            \\
                               & $ \Gamma_X=64.40$ $\gev$     & $
\Gamma_X=71.52$ $\gev$  &$\Gamma_X=92.84$ $\gev$   \\ 
\hline
Selection  cuts                & 231                         &   216  
                      &  111                      \\
$60<m_{ee} < 120$              & 231                         &   216  
                      &  111                       \\
$60 <m_{\mu \mu} < 120$        & 222                         &   208  
                      &  106                       \\
$1000 < m_{4\ell}$             & 221                         &   207  
                      &  106                        \\
\hline
\end{tabular}
\caption{Effects of the sequential cuts on the simulated events at $14~\tev$
3000 fb$^{-1}$ LHC for different values of $M_X$ and $\Gamma_X$ of a spin-$1^+$
resonance. It is easy to observe from the benchmark scenarios considered in this
table that at a given CM energy the production cross section decreases with
increase in $M_X$ and for a fixed value of $M_X$ the production cross section
decreases as the value of the decay width $\Gamma_X$ increases.}
\label{cut_table}
\end{table*}
So far we have not discussed about the background for the $p p \to X \to e^+ e^-
\mu^+ \mu^-$ process. This is discussed in the following subsection.

\subsection{Study of the angular asymmetries for a spin-\texorpdfstring{$1^{+}$}{1+} resonance:}\label{subsec:1p} 

\begin{figure}[hbtp!]
\centering \includegraphics[width=.9\linewidth]{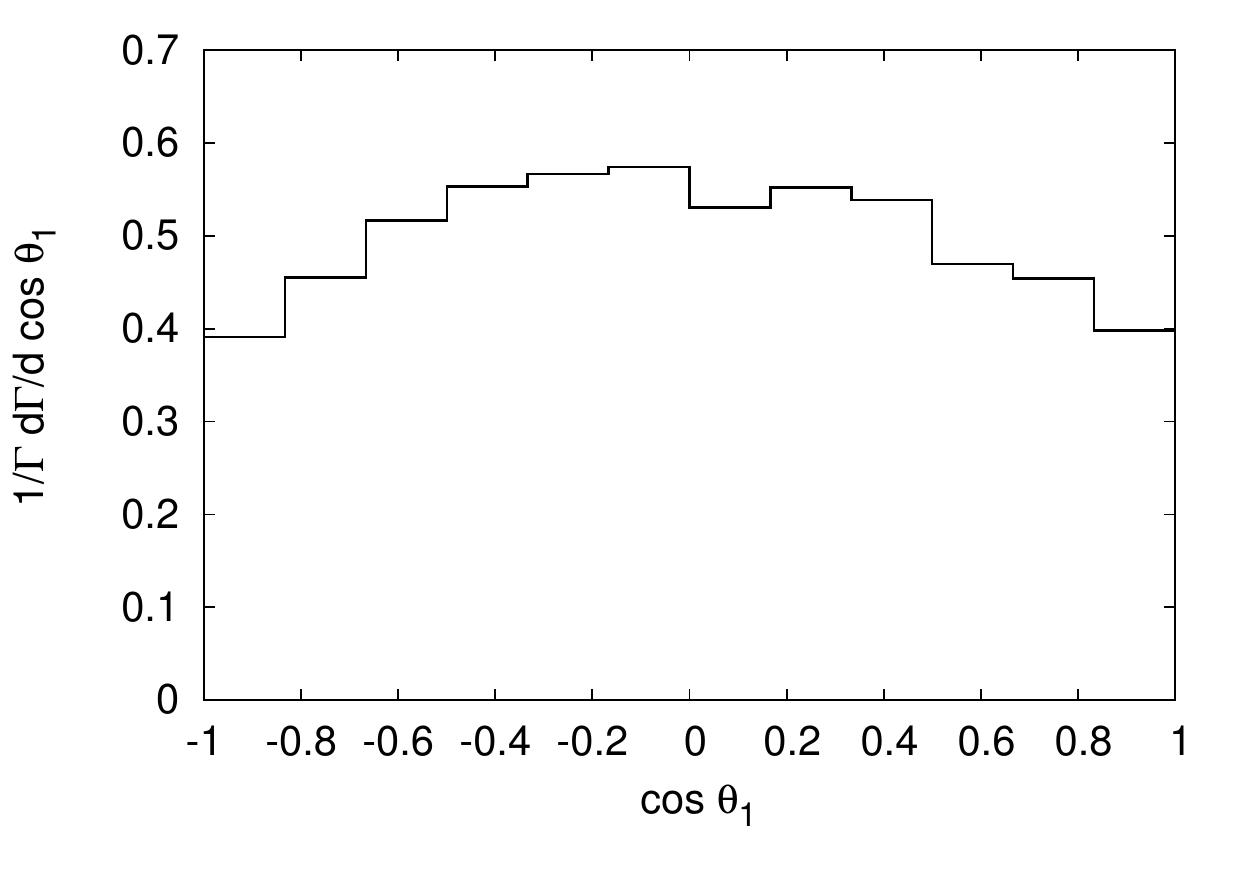} \caption{The
normalized distribution $\displaystyle \frac{1}{\Gamma}
\frac{d\Gamma}{d\cos\theta_1}$ vs. $\cos\theta_1$ for a Spin-$1^+$ resonance of
mass $M_X=1.8~\tev$ and width $\Gamma_X=64.04~\gev$.} \label{fig:ct1dist}
\end{figure}

\begin{figure}[hbtp!]
\centering \includegraphics[width=.9\linewidth]{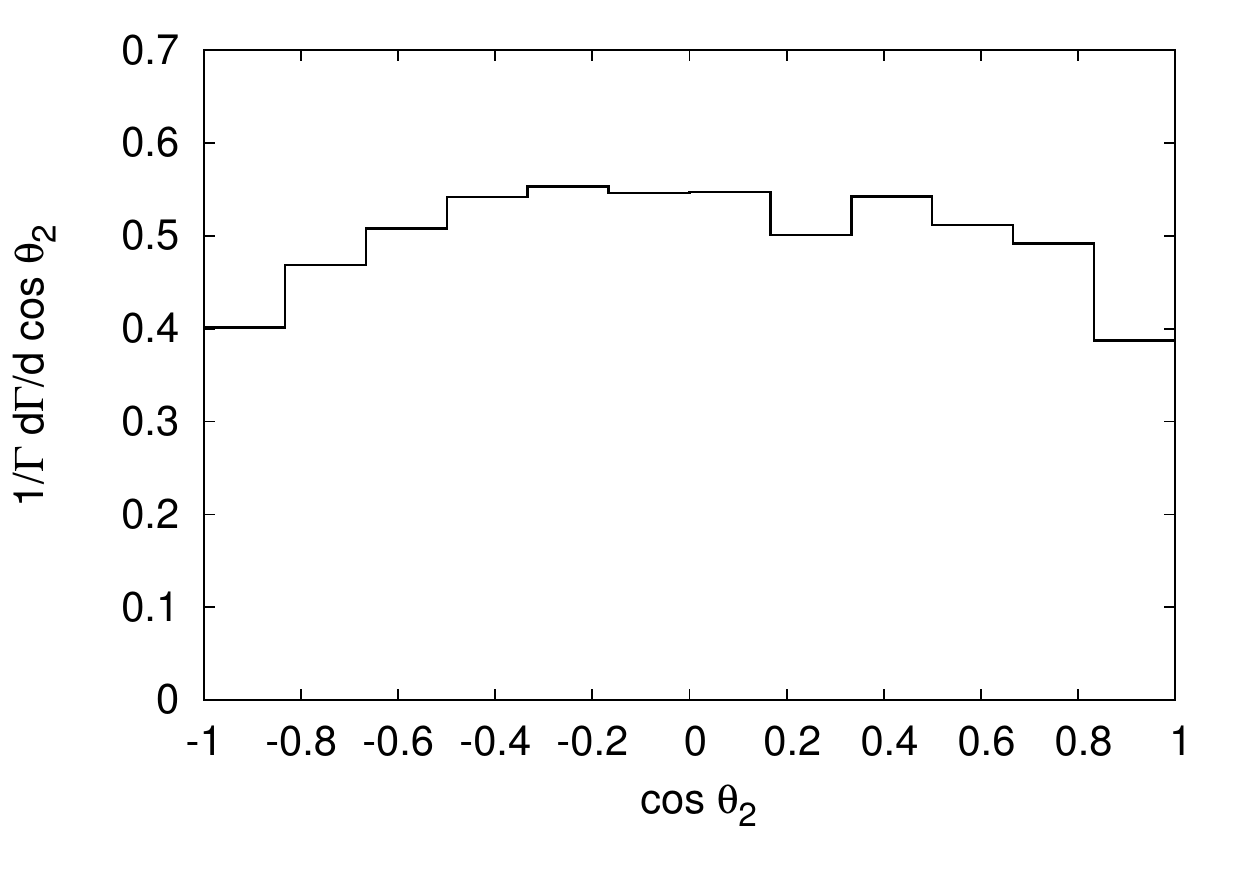} \caption{The
normalized distribution $\displaystyle \frac{1}{\Gamma}
\frac{d\Gamma}{d\cos\theta_2}$ vs. $\cos\theta_2$ for a Spin-$1^+$ resonance of
mass $M_X=1.8~\tev$ and width $\Gamma_X=64.04~\gev$. } \label{fig:ct2dist}
\end{figure}

\begin{figure}[hbtp!]
\centering \includegraphics[width=.9\linewidth]{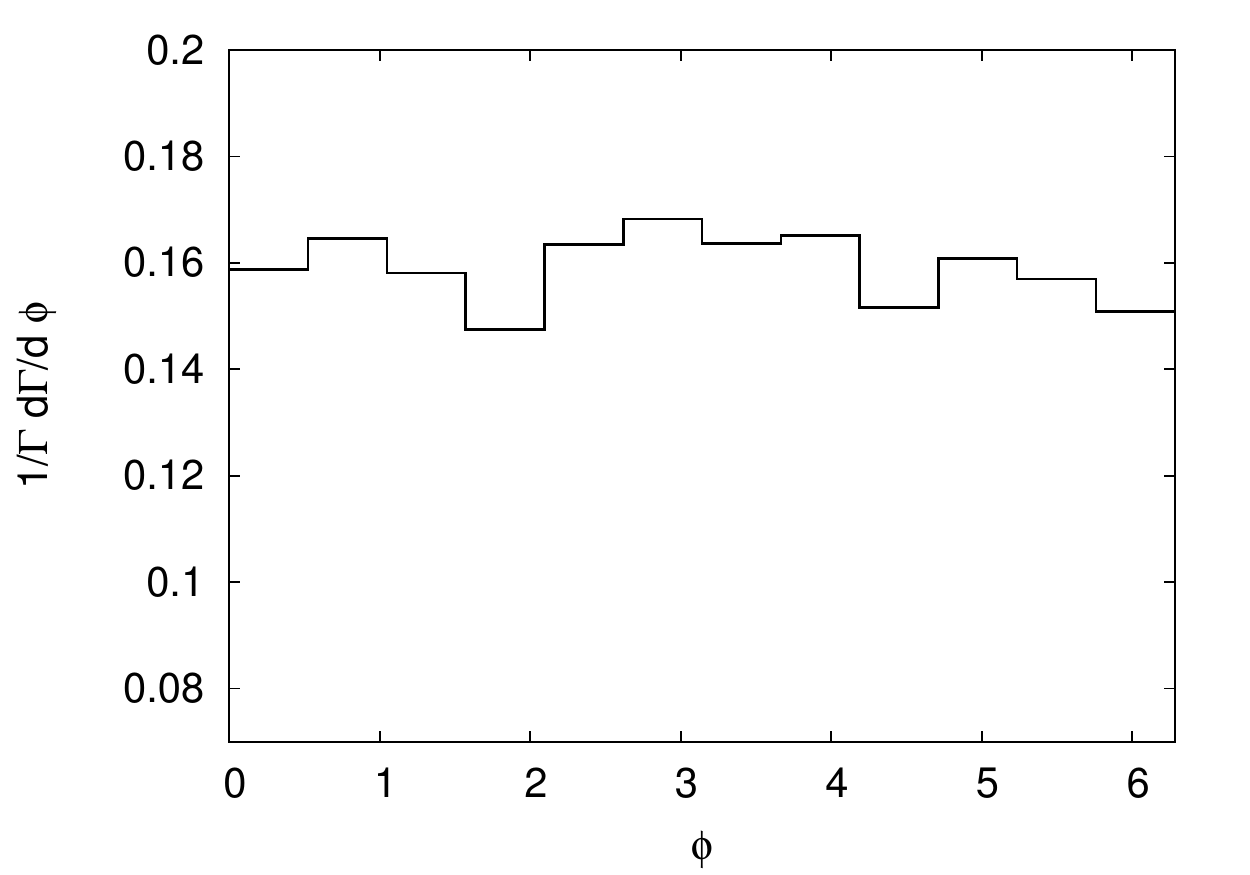} \caption{The normalized
distribution $\displaystyle \frac{1}{\Gamma} \frac{d\Gamma}{d\phi}$ vs. $\phi$
for a Spin-$1^+$ resonance of mass $M_X=1.8~\tev$ and width
$\Gamma_X=64.04~\gev$. } \label{fig:phidist}
\end{figure}

In this subsection we discuss the uniangular distributions and show how to
extract the angular observables from them for a spin-$1^+$ resonance of mass
$M_X=1.8~\tev$ and decay width $\Gamma_X= 64.40~\gev$. We choose this benchmark
scenario as the statistics is higher for this than the other cases. We analyse
the angular observables for this benchmark scenario for two different CM
energies $14~\tev$ and $33~\tev$ at an integrated luminosity 3000 fb$^{-1}$ in
future LHC runs.  The values of the couplings for this benchmark scenarios are
$c_q=0.12$ and $E_1=7.00\times 10^{-2}$.  The effects of the sequential cuts for
the benchmark scenarios are tabulated in Table~\ref{cut_tablep}.  The three
uniangular distributions for a spin-$1^+$ resonance, $\displaystyle
\frac{1}{\Gamma} \frac{d\Gamma}{d\cos\theta_1}$ vs. $\cos\theta_1$,
$\displaystyle \frac{1}{\Gamma} \frac{d\Gamma}{d\cos\theta_2}$ vs.
$\cos\theta_2$ and $\displaystyle \frac{1}{\Gamma} \frac{d\Gamma}{d\phi}$ vs.
$\phi$ are shown in Figs.~\ref{fig:ct1dist}, \ref{fig:ct2dist} and
\ref{fig:phidist} respectively.  It should be noted that the uniangular
distributions cover the full kinematic ranges for the three variables
$\cos\theta_1$, $\cos\theta_2$ and $\phi$.
\begin{table}[hbtp!]
 \centering
\begin{tabular}{|c|c|c|c|}
\hline
Cuts  in $\gev$                   & $14~\tev$,                  &  33
TeV,        \\
                                  & 3000 fb$^{-1}$           &  3000
fb$^{-1}$   \\ 
\hline
Selection  cuts                   &  231                     &   1212 
            \\
$60<m_{ee} < 120$                 &  231                     &   1212 
               \\
$60 <m_{\mu \mu} <120$            &  222                     &   1159 
               \\
$1000< m_{4\ell}$                 &  221                     &   1154 
                 \\
\hline
\end{tabular}
\caption{The effects of the sequential cuts on the simulated signal events at
$14~\tev$ and $33~\tev$ LHC with 3000 fb$^{-1}$ luminosity for a
spin-$1^{+}$resonance with $M_X= 1.8~\tev$ and width $\Gamma_X=64.40~ \gev$.}
\label{cut_tablep}
\end{table}

However, while extracting observables one has to take the background processes
into account. The $p p \to e^+ e^- \mu^+ \mu^-$ process is a continuum
background to the process $p p \to ZZ \to e^+ e^- \mu^+ \mu^-$. The effects of
the sequential cuts on the background process for $14~\tev$ and $33~\tev$ 3000
fb$^{-1}$ LHC run, are shown in the Table~\ref{cut_tablebkg}.
\begin{table}[hbtp!]
 \centering
\begin{tabular}{|c|c|c|c|}
\hline
Cuts  in $\gev$                   & $14~\tev$,                  &  33
TeV,        \\
                                                                      
                                  & 3000 fb$^{-1}$           &  3000
fb$^{-1}$   \\ 
\hline
Selection  cuts                   &  24530                   &  48588 
           \\
$60 < m_{ee} < 120$               &  23320                   &  46949 
               \\
$60 < m_{\mu \mu} <120$           &  18468                   &  40082 
              \\
$1000< m_{4\ell}$                 &  41                      &   238  
                \\
\hline
\end{tabular}
\caption{The effects of the sequential cuts on the simulated background events
at $14~\tev$ and $33~\tev$ LHC with 3000 fb$^{-1}$ luminosity.}
\label{cut_tablebkg}
\end{table}
The simulated signal and background events are finally binned in $\cos\theta_1$,
$\cos\theta_2$ and $\phi$ and fitted using Eqs.~\eqref{eq:cost1},
\eqref{eq:cost2} and \eqref{eq:phi} integrated over $m_{ee}^2 \left(\equiv q_1^2
\right)$ and $m_{\mu \mu}^2 \left( \equiv q_2^2 \right)$ to obtain the
\textit{integrated} angular observables with their respective errors.  The
values of the observables are tabulated in Table~\ref{observeablesp} for the two
different CM energies.
\begin{table}[hbtp!]
\centering
\begin{tabular}{c|c|c }
\hline
\hline
\textbf{Observables} & $14~\tev$, 3000 fb$^{-1}$                & 33
TeV, 3000  fb$^{-1}$ \\
\hline

$T_2$  &     $-0.19 \pm 0.11$                     & $-0.18\pm 0.06$   
  \\

$T_1$  &     $0.07\pm 0.09$                       & $0.01 \pm 0.04$\\

$T_2'$  &     $-0.09 \pm 0.12$                     & $-0.10 \pm 0.06$ 
    \\

$T_1'$  &     $-0.04 \pm 0.10$                     & $-0.03 \pm
0.05$\\

$U_2$             &     $0.08 \pm 0.51$                      & $0.04
\pm 0.24$ \\

$U_1$             &      $(-0.87\pm 5.33 )\times10^{-1}$     &
$(-0.17\pm 2.40 )\times10^{-1}$\\

$V_2$             &     $(-0.41\pm 5.32 )\times10^{-1}$      & 
$(0.21\pm 2.36 )\times10^{-1}$ \\

$V_1$             &     $(-0.32\pm 5.11 )\times10^{-1}$      & 
$(0.30\pm 2.34 )\times10^{-1}$ \\
\hline
\end{tabular}
\caption{The fit values and the respective errors of the observables $T_1$,
$T_2$, $U_1$, $U_2$, $V_1$ and $V_2$ for a Spin-$1^+$ resonance of mass
$M_X=1.8~\tev$ and width $\Gamma_X=64.04$ $\gev$ at $14~\tev$ and $33~\tev$ LHC
run (with 3000 fb$^{-1}$ luminosity).} \label{observeablesp}
\end{table}
It is clear from Table~\ref{observeablesp} that the observable $T_2$ and $T'_2$
extracted from $\cos \theta_1$ and $\cos \theta_2$ distributions respectively,
match within $1\sigma$ error. This is expected since both $q_1^2$ and $q_2^2$
are integrated over the same range. A full implementation of the flow chart
(shown in Fig.~\ref{fig:flowchart}) will require a fit with at least two regions
$q_1^2<q_2^2$ and $q_1^2>q_2^2$. However, given the heavy mass for $X$ the
production cross section is low, hence, the errors are still large and more
statistics is needed to undertake such a study.


\subsection{Study of the angular asymmetries for a spin-\texorpdfstring{$1^{-}$}{1-} resonance}\label{subsec:1m}

\begin{figure}[hbtp!]
\centering \includegraphics[width=.9\linewidth]{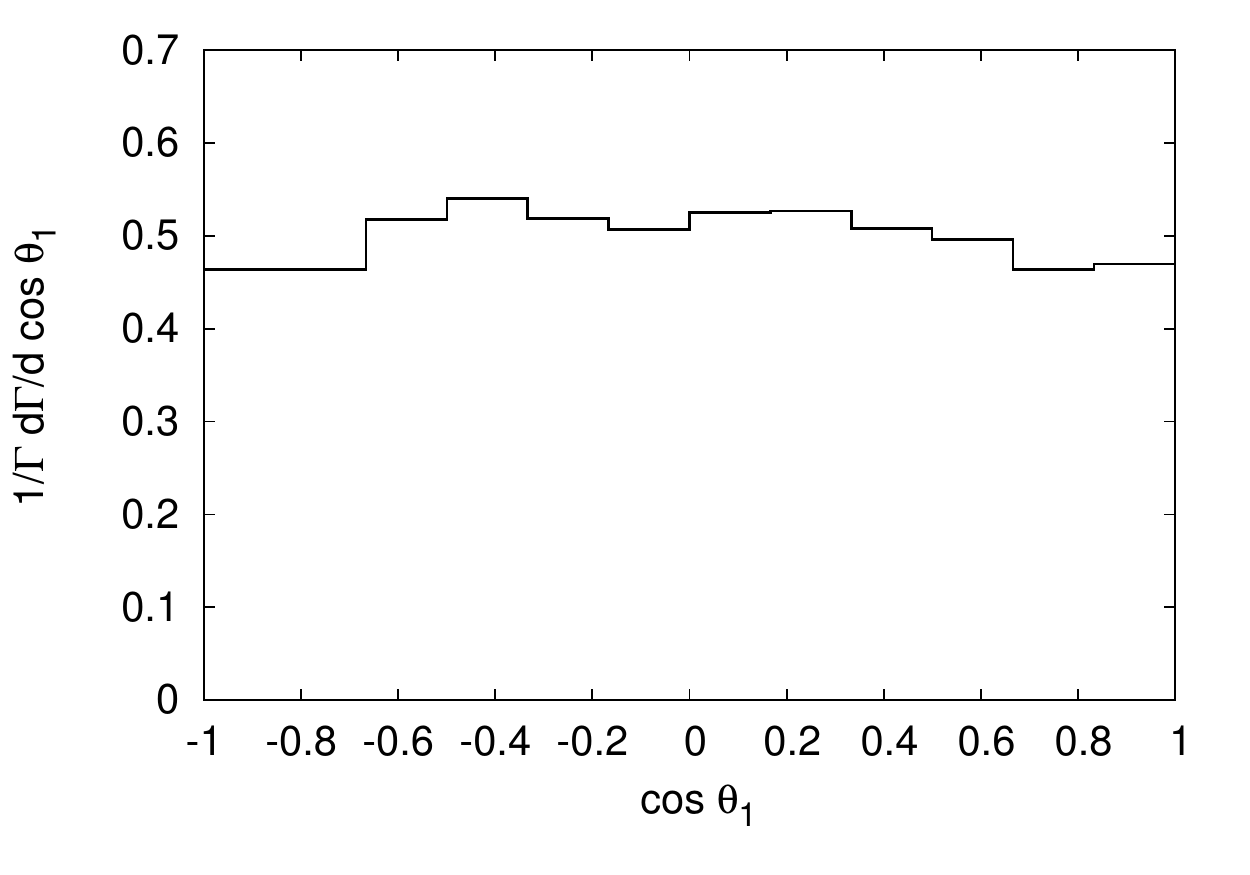} \caption{The
normalized distribution $\displaystyle \frac{1}{\Gamma}
\frac{d\Gamma}{d\cos\theta_1}$ vs.  $\cos\theta_1$ for a Spin 1${^-}$ resonance
of mass $M_X=1.8~\tev$ and width $\Gamma_X=64.04$ $\gev$.} \label{fig:ct1distm}
\end{figure}

\begin{figure}[hbtp!]
\centering \includegraphics[width=.9\linewidth]{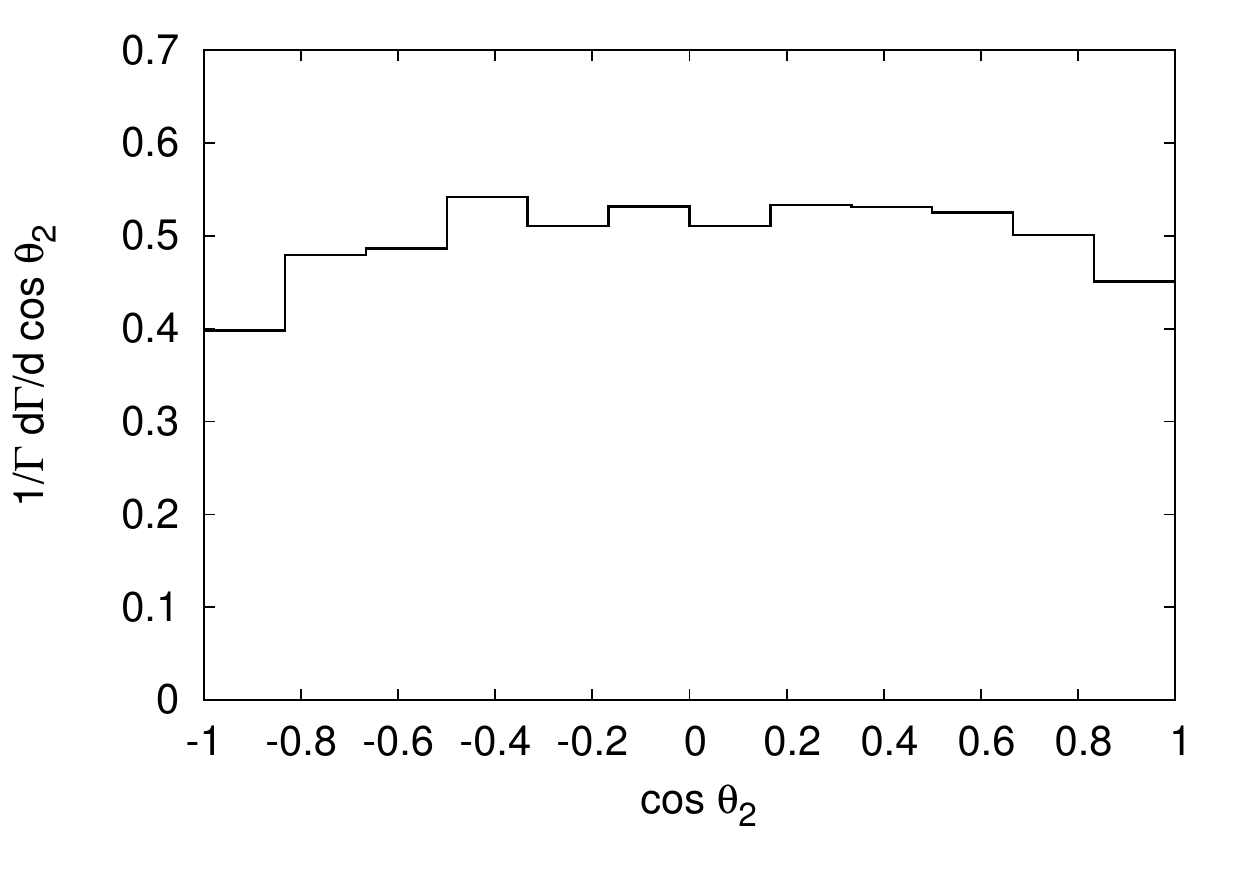} \caption{The
normalized distribution $\displaystyle \frac{1}{\Gamma}
\frac{d\Gamma}{d\cos\theta_2}$ vs. $\cos\theta_2$ for a Spin 1${^-}$ resonance
of mass $M_X=1.8~\tev$ and width ~$\Gamma_X=64.04$ $\gev$. }
\label{fig:ct2distm}
\end{figure}

\begin{figure}[hbtp!]
\centering \includegraphics[width=.9\linewidth]{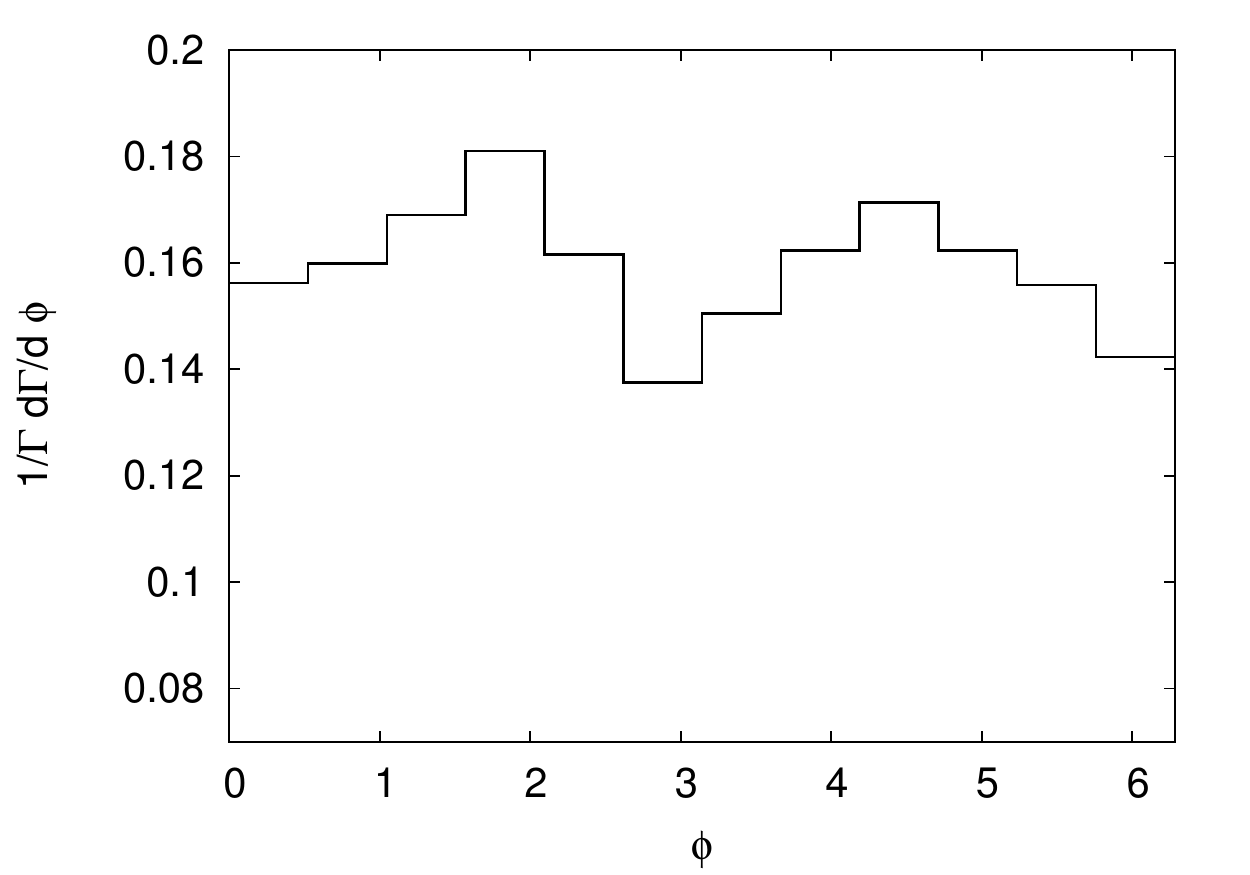} \caption{The
normalized distribution $\displaystyle \frac{1}{\Gamma} \frac{d\Gamma}{d\phi}$
vs. $\phi$ for a Spin 1${^-}$ resonance of mass $M_X=1.8~\tev$ and width
$\Gamma_X=64.04$ $\gev$. } \label{fig:phidistm}
\end{figure}

We have so far discussed the possibility of finding a heavy spin-$1^{+}$
resonance. However, the resonance may well could be a spin-$1^-$.  The limits on
the couplings $\tilde{c}_q$ and $O_1$ can also be found from $\sigma\times Br
\times \mathcal{A}$ limit from Ref.~\cite{Khachatryan:2015sja}.  In the limit
$M_Z \ll M_X$ and $m_q \ll M_X$, the couplings $\tilde{c}_q\approx c_q$ and
$O_1\approx E_1$. Hence, we choose the values $\tilde{c}_q=0.12$ and
$O_1=7.00\times 10^{-2}$ for the couplings as a benchmark scenario for our
analysis of a spin-$1^-$ resonance of mass $1.8~\tev$ and decay width
$\Gamma_X=64.40~\gev$. We perform the same analysis as given in
Sec.~\ref{subsec:1p} and extract out the values of the angular observables at
two different CM energies, $14~\tev$ and $33~\tev$ at an integrated luminosity
of 3000 fb$^{-1}$. We also find three uniangualr distributions for the
spin-$1^-$ resonance, shown in Figs.~\ref{fig:ct1distm}, \ref{fig:ct2distm} and
\ref{fig:phidistm} respectively.  The effects of the sequential cuts are
tabulated in Table~\ref{cut_tablem} at $14~\tev$ and $33~\tev$ LHC for 3000
fb$^{-1}$ luminosity.  The background analysis for spin-$1^-$ resonance would
remain the same as stated in Sec~\ref{subsec:1p}.
\begin{table}[hbtp!]
 \centering
\begin{tabular}{|c|c|c|c|}
\hline
Cuts  in $\gev$               & $14~\tev$,                  & 
$33~\tev$,       \\
                              & 3000 fb$^{-1}$            &  3000
fb$^{-1}$   \\ 
\hline
Selection  cuts               &  220                     &   1155     
        \\
$60<m_{ee} < 120$             &  200                     &   1154     
           \\
$60<m_{\mu \mu} < 120$             &  211                     &   1108
                \\
$1000< m_{4\ell}$             &  210                     &   1105     
             \\
\hline
\end{tabular}
\caption{The effects of the sequential cuts on the simulated signal events at
$14~\tev$ 3000 fb$^{-1}$ and $33~\tev$ 3000 fb$^{-1}$ LHC of a
spin-$1^{-}$resonance with $M_X=1.8~\tev$ and width $\Gamma_X=64.40$ $\gev$.}
\label{cut_tablem}
\end{table}

The observables extracted from the uniangular distributions of the
spin-$1^-$resonance are given in Table~\ref{observeablesm}.
\begin{table}[hbtp!]
\centering
\begin{tabular}{c|c|c }
\hline
\hline
\textbf{Observables} & $14~\tev$, 3000 fb$^{-1}$              & 33
TeV, 3000  fb$^{-1}$ \\
\hline

$T_2$  &     $-0.07\pm 0.13$                    &  $-0.12\pm 0.06$      \\

$T_1$  &     $0.06 \pm 0.10$                    &  $0.01 \pm 0.04$\\

$T_2'$  &     $-0.04 \pm 0.13$                    &  $-0.07 \pm 0.06$      \\

$T_1'$  &     $ 0.09 \pm 0.10$                    &  $-0.01 \pm 0.04$\\

$U_2$             &     $-0.08 \pm 0.54$                    &  $-0.05 \pm 0.24$ \\

$U_1$             &     $(1.99\pm 5.16 )\times10^{-1}$     &  $(1.20\pm 2.32 )\times10^{-1}$\\

$V_2$             &     $(-0.14\pm 5.27 )\times10^{-1}$     &  $(0.58\pm 2.33 )\times10^{-1}$ \\

$V_1$             &     $(-0.27\pm 5.46 )\times10^{-1}$      &  $(-0.63\pm 2.39 )\times10^{-1}$ \\
\hline
\end{tabular}
\caption{The fit values and the errors of the observables $T_1$, $T_2$, $U_1$,
$U_2$, $V_1$ and $V_2$ for a Spin 1${^-}$ resonance of mass $M_X=1.8~\tev$ and
width $\Gamma_X=64.04$ $\gev$ for signal plus background. The values are
extracted for two different CM energies, $14~\tev$ and $33~\tev$ LHC run with
luminosity 3000 fb$^{-1}$.} \label{observeablesm}
\end{table}
Apart from $T_2$ the errors of the other observables are still not small and
require higher statistics to fully study the flowchart \ref{fig:flowchart}.

If a heavy spin-$1$ resonance is seen at the LHC, the full angular analysis and
the extraction of all the observables may not be entirely possible at $33~\tev$
3000 fb$^{-1}$ run.  Once such resonance is observed a future high luminous
machine could disentangle the exact spin and parity of the resonance studying
the observables extracted from uniangular distributions.  Moreover, we have not
discussed the spin-$1$ resonance with mixed parity configuration and would
typically require higher statistics as well, to completely disentangle, which is
beyond the scope of the current paper.

\section{Conclusion} \label{sec:conclusion} 

We conclude that by looking at three \textit{normalized} uni-angular
distributions, for the decay of a resonance to four charged final leptons via
two $Z$ bosons, one can infer to a fairly good accuracy the spin and parity of
the parent particle. We show that it is possible for a \textit{special} $2^{+}$
resonance to give angular distributions comparable to those of a $0^{+}$
resonance. It is in this special case that one needs to study the $Y^2$
dependence of the helicity amplitudes in order to distinguish the two cases.
Since the spin-$1$ case has only two helicity amplitudes, it needs minimum
number of observables to get confirmed or ruled out. We have also provided a
step-by-step methodology that must be followed to distinguish the various spin,
parity possibilities that are allowed in the case under consideration.  A
numerical analysis has also been performed, for a heavy spin $1$ resonance to
validate our formalism. It would therefore be not an overstatement to say that
this method can play a crucial role, at future high luminosity machines, in
discovering the spin-parity nature of any new resonance, such as a heavy scalar
boson or a $Z'$ boson or a Kaluza-Klein boson or any such resonance, found to
decay to four final charged leptons via two $Z$ bosons.

\begin{acknowledgements}
DS is thankful to HYC and Institute of Physics, Academia Sinica, Taiwan, R.O.C.\
for hospitality. HYC and TCY were supported in part by Ministry of Science and
Technology of R.O.C.\ under the Grant Nos.~103-2112-M-001-005 and
101-2112-M-001-005-MY3 respectively.  We thank Prof.\ Arjun Menon for many
fruitful discussions on this work.
\end{acknowledgements}

\appendix

\section{Relationships amongst the form factors and the effective coupling constants}
\label{sec:appendix-form-factors-coupling-constants}

The form factors $\Even{i}{J}, \Odd{i}{J}$ (which enter the vertex factors in
Eqs.~\eqref{eq:vertex-0}, \eqref{eq:vertex-1} and \eqref{eq:vertex-2}) are
related to the effective coupling constants $\even{i}{J}, \odd{i}{J}$ (which
eneter the Lagrangians given in Eqs.~\eqref{eq:Lagrangian-0},
\eqref{eq:Lagrangian-1} and \eqref{eq:Lagrangian-2}) in the following manner
\begin{align}
\Even{1}{0} &= \even{1}{0} + 4 \, \even{2}{0} \, (q_1 \cdot q_2), \\
\Even{2}{0} &= - 4 \, \even{2}{0}, \\
\Odd{1}{0} &= - 8 \, \odd{1}{0}, \\
\Odd{1}{1} &= - \odd{1}{1}, \\
\Even{1}{1} &= 2 \even{1}{1}, \\
\Even{1}{2} &= \even{1}{2} + \frac{1}{4} \even{2}{2} (P^2-Q^2) +
\frac{1}{16} \even{3}{2} (P^2 - Q^2)^2 \nn \\
 & \qquad + \frac{1}{4} \even{4}{2} (P^4-(P \cdot Q)^2),\\
\Even{2}{2} &= \frac{1}{4} \Big( 4 \even{2}{2} + \even{3}{2} (P^2 -
Q^2) \nn\\
 & \qquad \qquad + 2 \even{4}{2} (P^2 + P \cdot Q) \Big),\\
\Even{3}{2} &= \frac{1}{4} \left( - 2 \even{2}{2} - \even{5}{2} (P^2 -
Q^2) \right),\\
\Even{4}{2} &= \frac{1}{2} \left( \even{3}{2} + \even{4}{2} - 2
\even{5}{2} \right),\\
\Odd{1}{2} &= 4 \odd{1}{2} + \odd{2}{2} (P^2 - Q^2) + 2 \odd{3}{2}
(P^2 + P \cdot Q),\\
\Odd{2}{2} &= \odd{2}{2} + \odd{3}{2},\\
\Odd{3}{2} &= - \odd{1}{2},\\
\Odd{4}{2} &= \odd{4}{2}.
\end{align}

\section{Helicity amplitudes and partial wave contributions for the decay \texorpdfstring{$X \to ZZ$}{X->ZZ}}
\label{sec:helicity-partial-wave} 

If we specify the polarizations of the initial and final particles, then the
Feynman amplitude or transition amplitude can always be written in terms of
helicity amplitudes. We shall represent the polarisation state of a particle by
a ket $\ket{\mbox{spin}, \mbox{spin projection to } z \mbox{ axis}}$. Then the
Feynman amplitude for the
process $$\underbrace{\ket{\mathbf{J},J_z}}_{X} \to
\underbrace{\ket{\mathbf{1},\lambda_1}}_{Z_1} \;
\underbrace{\ket{\mathbf{1},\lambda_2}}_{Z_2}$$ is given by the well known
expression \cite{Jacob:1959at,Chung:1102240} involving the Wigner-$D$
function $\mathscr{D}_{J_z \lambda}^{J*} (\phi, \theta, -\phi)$:
\begin{align}
  \mathscr{M}(J_z, \lambda_1, \lambda_2) &= \left( \frac{2J + 1}{4\pi}
  \right)^{\frac{1}{2}} \mathscr{D}_{J_z
    \lambda}^{J*}(\phi,\theta,-\phi) \; A_{\lambda_1 \lambda_2},
\end{align}
where $\lambda = \modulus{\lambda_1 - \lambda_2}$ with $\lambda_{1,2} \in \{ \pm
1, 0 \}$, $J = \modulus{\mathbf{J}}$, and $A_{\lambda_1 \lambda_2}$ is called
the \textit{helicity amplitude}. Conservation of angular momentum implies that
\begin{equation}
  \label{eq:lambda}
  \modulus{\lambda} = \modulus{\lambda_1 - \lambda_2} \leqslant J.
\end{equation}
Since there are no interferences amongst the amplitudes with different helicity
configurations, we will have to sum over all the allowed values of $\lambda_1$
and $\lambda_2$ that are not constrained by the value of $J_z$ after squaring
each individual amplitude:
\begin{align}
  \modulus{\mathscr{M}}^2 &= \sum_{\substack{\lambda_1, \lambda_2 \\
      \modulus{\lambda_1 - \lambda_2} \leqslant J}}
  \modulus{\mathscr{M} \left( J_z, \lambda_1, \lambda_2 \right)}^2
  \nn\\
  &= \left( \frac{2 J + 1}{4\pi} \right) \sum_{\substack{\lambda_1,
      \lambda_2 \\
\modulus{\lambda_1 - \lambda_2} \leqslant J}}
  \modulus{\mathscr{D}^{J*}_{J_z \lambda} \left( \phi, \theta, -\phi
    \right)}^2 \; \modulus{A_{\lambda_1 \lambda_2}}^2 .
\end{align}
Thus the probability of contribution of the helicity amplitude $A_{\lambda_1
\lambda_2}$ to the transition amplitude ca be found as $\mathscr{M} \left( J_z,
\lambda_1, \lambda_2 \right)$ is $\displaystyle \left( \frac{2 J + 1}{4\pi}
\right) \modulus{\mathscr{D}^{J*}_{J_z \lambda} \left( \phi, \theta, -\phi
\right)}^2$. We can therefore write down the following important fact of the
helicity amplitude formalism: \textit{All the allowed helicity amplitudes for a
given decay process contribute, but with different definite probability, to the
Feynman amplitude, irrespective of the polarization of the parent (decaying)
particle. The probability, however, depends on the polarization of the parent
particle and for all allowed helicity amplitudes is non-zero.} Since the two $Z$
bosons are Bose symmetric, the helicity amplitudes satisfy the relation
\begin{equation}
  \label{eq:helicity-bose-symmetry}
  A_{\lambda_2 \lambda_1} = (-1)^J A_{\lambda_1 \lambda_2} =
  \begin{cases}
    +A_{\lambda_1 \lambda_2} & \mbox{ for spin-}0,2\\
    -A_{\lambda_1 \lambda_2} & \mbox{ for spin-}1
  \end{cases}.
\end{equation}
This relationship is useful in getting the correct number of independent
helicity amplitudes. All the allowed helicity amplitudes in the decay $X \to ZZ$
are given in Table~\ref{tab:helicity-amplitudes-spin} where N denotes the total
number independent helicity amplitudes possible for the particular spin case.

\begin{table}[hbtp]
\centering
\begin{tabular}[c]{|c|l|c|}\hline
  Spin of $X$ & Allowed Helicity Amplitudes & N \\ \hline
  $0$ & $A_{++}$, $A_{00}$, $A_{--}$. & 3 \\ \hline
  $1$ & $A_{+0} = - A_{0+}$, $A_{0-}= - A_{-0}$. & 2 \\ \hline
  \multicolumn{1}{|c|}{\multirow{2}{*}{$2$}} & $A_{++}$, $A_{00}$,
  $A_{--}$, $A_{+-} = A_{-+}$, &
\multicolumn{1}{c|}{\multirow{2}{*}{6}} \\
  & $A_{+0} = A_{0+}$, $A_{0-}= A_{-0}$. & \\ \hline
\end{tabular}
\caption{Allowed helicity amplitudes considering only the different spin
possibilities.} \label{tab:helicity-amplitudes-spin}
\end{table}

It is also known that, if the particle $X$ were a parity eigenstate with
eigenvalue $\eta_X = + 1 $ (parity-even) or $-1$ (parity-odd), then the helicity
amplitudes are related by:
\begin{equation}
 A_{\lambda_1 \lambda_2} = \eta_X \, \left(-1\right)^J \,
 A_{-\lambda_1 \, -\lambda_2}.
\end{equation}
The allowed helicity amplitudes for the different spin-parity possibilities can
thus be related and are given in
Table~\ref{tab:helicity-amplitudes-spin-parity}.
\begin{table}[hbtp]
\centering
\begin{tabular}[c]{|c|l|c|} \hline
$J^P$ of $X$ & Allowed Helicity Amplitudes & N \\ \hline
$0^+$ & $A_{++} = A_{--}$, $A_{00}$. & 2 \\ \hline 
$0^-$ & $A_{++} = - A_{--}$. & 1 \\ \hline
$1^+$ & $A_{+0} = - A_{-0} = A_{0-} = - A_{0+}$. & 1 \\ \hline 
$1^-$ & $A_{+0} = A_{-0} = - A_{0-} = - A_{0+}$. & 1 \\ \hline 
\multicolumn{1}{|c|}{\multirow{2}{*}{$2^+$}} & $A_{++} = A_{--}$,
$A_{00}$, $A_{+-} = A_{-+}$, &
\multicolumn{1}{c|}{\multirow{2}{*}{4}} \\
 & $A_{+0} = A_{-0} = A_{0-} = A_{0+}$. & \\ \hline 
\multicolumn{1}{|c|}{\multirow{2}{*}{$2^{-}$}} & $A_{++} = - A_{--}$,
& \multicolumn{1}{c|}{\multirow{2}{*}{2}} \\
 & $A_{+0} = - A_{-0} = - A_{0-} = A_{0+}$. & \\ \hline 
\end{tabular}
\caption{Relationships amongst the allowed helicity amplitudes for the different
spin-parity cases.} \label{tab:helicity-amplitudes-spin-parity}
\end{table}
It is clearly evident from above that for the spin-0 case out of the three
helicity amplitudes two describe the parity-even scenario and only one describes
the parity-odd scenario. Similarly, for spin-1 both parity-even and parity-odd
cases are described by one helicity amplitude each. For spin-2 case, we have
four helicity amplitudes describing the parity-even scenario and two helicity
amplitudes for the parity-odd scenario.

If we now make a change of basis from the helicity basis to the transversity
basis (also called as linear polarization basis), then the number of independent
transversity amplitudes must be equal to the number of helicity amplitudes, and
\textit{all the allowed transversity amplitudes would contribute with definite
probability to the Feynman amplitude irrespective of the polarization of the
parent (decaying) particle.}

Let us now analyse the decay process from the point-of-view of partial wave
decompositions. If we describe the two $Z$ boson system by a ket specifying the
total spin ($\mathbf{L}_{\text{spin}}$), the relative orbital angular momentum
($\mathbf{L}_{\text{orbital}}$), the spin of the parent particle (its
$\mathbf{J}$ here) and its projection along the direction of flight of one of
the $Z$ bosons ($J_z$):
$\ket{\mathbf{J},J_z;\mathbf{L}_{\text{orbital}},\mathbf{L}_{\text{ spin }}}$,
then
\begin{align}
  & \hat{P}_{12} \ket{\mathbf{J},J_z ; \mathbf{L}_{\text{orbital}},
    \mathbf{L}_{\text{spin}}} \nn\\
  & = (-1)^{L_{\text{orbital}} + L_{\text{spin}}} \ket{\mathbf{J},J_z
    ; \mathbf{L}_{\text{orbital}}, \mathbf{L}_{\text{spin}}},
\end{align}
where $\hat{P}_{12}$ is the operator that exchanges the two $Z$ bosons (it
exchanges both their momenta and spins or polarisations), $L_{\text{orbital}}$
and $L_{\text{spin}}$ are the modulus of $\mathbf{L}_{\text{orbital}}$ and
$\mathbf{L}_{\text{spin}}$ respectively.  It is obvious that for Bose symmetry
to be satisfied $L_{\text{orbital}} + L_{\text{spin}}$ must be even. The allowed
partial waves for the decay $X \to ZZ$ are enunciated in
Table.~\ref{tab:allowed-partial-waves}.

\begin{table}[hbtp]
\centering
\begin{tabular}[c]{|c|c|c|c|} \hline
$\mathbf{L}_{\text{spin}}$ & $\mathbf{L}_{\text{orbital}}$ &
  $\mathbf{J}$ & Partial wave \\ \hline
$\mathbf{0}$ & $\mathbf{0}$ & $\mathbf{0}$ & $S$-wave \\ \hline
$\mathbf{0}$ & $\mathbf{2}$ & $\mathbf{2}$ & $D$-wave \\ \hline
$\mathbf{1}$ & $\mathbf{1}$ & $\mathbf{2}, \mathbf{1}, \mathbf{0}$ &
$P$-wave \\ \hline
$\mathbf{1}$ & $\mathbf{3}$ & $\mathbf{2}$ & $F$-wave \\ \hline
$\mathbf{2}$ & $\mathbf{0}$ & $\mathbf{2}$ & $S$-wave \\ \hline
$\mathbf{2}$ & $\mathbf{2}$ & $\mathbf{2}, \mathbf{1}, \mathbf{0}$ &
$D$-wave \\ \hline
$\mathbf{2}$ & $\mathbf{4}$ & $\mathbf{2}$ & $G$-wave \\ \hline
\end{tabular}
\caption{Allowed partial waves for all the spin considerations.}
\label{tab:allowed-partial-waves}
\end{table}

It is easy to observe that when $X$ has spin-$0$, then there are three helicity
amplitudes and three partial wave contributions (one $S$-wave, one $P$-wave and
one $D$-wave). When $X$ has spin-$1$, then there are only two independent
helicity amplitudes and two partial wave contributions (one $P$-wave and one
$D$-wave). Finally when $X$ has spin-$2$, then there are six independent
helicity amplitudes and six partial wave contributions (one $S$-wave, one
$P$-wave, two $D$-waves, one $F$-wave and one $G$-wave). It is interesting to
note that for the spin-$0$ case the vertex factor has three form factors, for
spin-$1$ case there are two form factors. However, for the spin-$2$ case we have
eight form factors in the vertex factor instead of six. So one needs to consider
only six form factors out of which four should be of parity-even nature and two
should be of parity-odd nature. Out of the four parity-odd form factors, only
$\Odd{1}{2}$ and $\Odd{2}{2}$ contribute to the vertex factor as explained in
\ref{sec:appendix-redundancy-in-spin-2-vertex}.

\section{Redundancy of inclusion of \texorpdfstring{$\Odd{3}{2}, \Odd{4}{2}$}{O3(2), O4(2)} in the spin-2 vertex factor}
\label{sec:appendix-redundancy-in-spin-2-vertex}

If we include the $\Odd{3}{2}$ term in the vertex factor for spin-$2$ case, then
only the two helicity amplitudes $\Ampo{1}{2}$ and $\Ampo{2}{2}$ get modified as
follows:
\begin{align}
  \Ampo{1}{2} &= \frac{4 Y}{3 M_{+}} \bigg( \left(\Odd{1}{2} -
  \Odd{3}{2}\right) \left( M_{-}^4 - M_X^2 M_{+}^2 \right) \nn\\
  & \qquad \qquad \qquad \qquad + \Odd{2}{2} \left( 4 M_{+}^2 M_X^2
  Y^2 \right) \bigg),\\
  \Ampo{2}{2} &= \frac{8 M_1 M_2 \mu^2 Y}{3 \sqrt{3} M_{+}} \;
  \left(\Odd{1}{2} - \Odd{3}{2}\right),
\end{align}
where the previous $O_{1}^{(2)}$ is now replaced by $\left(\Odd{1}{2} -
\Odd{3}{2}\right)$. Note that only this combination of the two form factors
would be accessible to any experiment. Moreover, all other helicity amplitudes
remain unchanged. In the vertex factor of Eq.~\eqref{eq:vertex-2} the term which
is proportional to $\Odd{4}{2}$ can be rewritten using Schouten identity as
follows:
\begin{align}
& \epsilon^{\alpha \beta \rho \sigma} \; Q^{\mu} Q^{\nu} q_{1\rho}
q_{2\sigma} \nn\\
&= \frac{1}{2} \Big[ Q^{\nu} \left( \epsilon^{\alpha
\mu \rho \sigma} \; Q^{\beta} - \epsilon^{\beta \mu \rho \sigma} \;
Q^{\alpha} \right) \nn\\
& \qquad \qquad + Q^{\mu} \left( \epsilon^{\alpha \nu \rho
\sigma} \; Q^{\beta} - \epsilon^{\beta \nu \rho \sigma} \; Q^{\alpha}
\right) \Big] q_{1\rho} q_{2\sigma} \nn\\
& \quad  - \frac{Q^2}{4} \left( \epsilon^{\alpha \beta \mu \rho} \;
P_{\rho} Q^{\nu} + \epsilon^{\alpha \beta \nu \rho} \; P_{\rho}
Q^{\mu} \right) \nn\\
& \quad + \frac{P \cdot Q}{4} \left( \epsilon^{\alpha
\beta \mu \sigma} \; Q^{\nu} Q_{\sigma} + \epsilon^{\alpha \beta \nu
\sigma} \; Q^{\mu} Q_{\sigma} \right).
\end{align}
Thus the form factor $\Odd{4}{2}$ can be absorbed into the existing form factors
$\Odd{2}{2}$ and $\Odd{3}{2}$ by redefining them as follows:
\begin{align}
\Odd{2}{2} &\to \Odd{2}{2} + \frac{1}{2} \Odd{4}{2},\\
\Odd{3}{2} &\to \Odd{3}{2} - \frac{Q^2}{4} \Odd{4}{2}.
\end{align}
The remaining contribution of $\Odd{4}{2}$ is proportional to $P \cdot Q$. Since
$P \cdot Q$ is not Bose symmetric, this term does not contribute to our vertex
factor. Therefore the effect of including the $\Odd{3}{2}$ and $\Odd{4}{2}$
terms in the vertex factor can simply be taken care of by the following
redefinitions of $\Odd{1}{2}$ and $\Odd{2}{2}$ as follows:
\begin{align}
\Odd{1}{2} &\to \Odd{1}{2} - \Odd{3}{2} + \frac{Q^2}{4} \Odd{4}{2},\\
\Odd{2}{2} &\to \Odd{2}{2} + \frac{1}{2} \Odd{4}{2}.
\end{align}
Since we cannot have more than six helicity amplitudes for the spin-$2$ case,
and we already have six form factors in the vertex factor, any additional form
factor that we introduce in the vertex factor must come in association with the
existing form factors, as proved here.

\section{Expressions for \texorpdfstring{$\Ti{J}$, $\Tip{J}$, $\Ui{J}$, $\Vi{J}$}{T, T', U, V}} \label{sec:apendix-TTUV}

The expressions for the coefficients $\Ti{J}$, $\Tip{J}$, $\Ui{J}$, $\Vi{J}$
that entered the uni-angular distributions are given as follows:
\begin{align}
  \Tone{0} &= - \Tonep{0} = -\frac{3}{2} \, \eta \; \Re\left(\Fe{2}{0}
  \Fost{1}{0}\right),\\
  \Ttwo{0} &= \Ttwop{0} =\frac{1}{4}
  \left(1-3\modulus{\Fe{1}{0}}^2\right),\\
  \Uone{0} &=-\frac{9\pi^2}{32\sqrt{2}} \eta^2 \, \Re\left(\Fe{1}{0}
  \Fest{2}{0}\right),\\
  \Utwo{0} &=\frac{1}{4} \left(\modulus{\Fe{2}{0}}^2 -
  \modulus{\Fo{1}{0}}^2\right),\\
  \Vone{0} &=-\frac{9\pi^2}{32\sqrt{2}} \eta^2 \, \Im\left( \Fe{1}{0}
  \Fost{1}{0} \right),\\
  \Vtwo{0} &=\frac{1}{2} \, \Im\left(\Fe{2}{0} \Fost{1}{0} \right),\\
    \Tone{1} &= -\frac{6 \sqrt{2} \eta M_X^3 M_1^2 Y}{D_1 D_2}
  \left(M_X^2 - M_1^2 + 3 M_2^2\right) \; \Re\left( \Fe{1}{1}
  \Fost{1}{1} \right),\\
  \Tonep{1} &= \frac{6 \sqrt{2} \eta M_X^3 M_2^2 Y}{D_1 D_2}
  \left(M_X^2 + 3 M_1^2 - M_2^2\right) \; \Re\left( \Fe{1}{1}
  \Fost{1}{1} \right),\\
  \Ttwo{1} &= - 2 M_X^2 Y^2 \left( \frac{\modulus{\Fo{1}{1}}^2}{D_1^2}
  \bigg( \left(M_X^2 - M_1^2 \right) \left(M_1^2+4 M_2^2\right) + 2
  M_2^4 \bigg) \right. \nn\\
  & + \left. \frac{2 \modulus{\Fe{1}{1}}^2}{D_2^2} \bigg( M_X^2
    \left(M_1^2+16 M_2^2\right) - M_2^2 \left( 20 M_1^2 - 3 M_2^2
    \right) \bigg) \right), \\
  \Ttwop{1} &= - 2 M_X^2 Y^2 \left(
  \frac{\modulus{\Fo{1}{1}}^2}{D_1^2} \right.  \bigg( \left(M_X^2 -
  M_2^2 \right) \left(4 M_1^2+M_2^2\right) +2 M_1^4 \bigg) \nn\\
  & + \left. \frac{2 \modulus{\Fe{1}{1}}^2}{D_2^2} \bigg( M_X^2
  \left(16 M_1^2+M_2^2\right) - M_1^2 \left( 20 M_2^2 - 3 M_1^2
  \right) \bigg) \right),\\
  \Uone{1} &= \frac{9 \pi ^2 \eta ^2 M_X^2 M_1 M_2}{16 D_1^2 D_2^2}
  \bigg( 4 \modulus{\Fo{1}{1}}^2 M_X^2 Y^2 \Big( 16 M_X^6
  M_{+}^2 \nn\\
  &+M_X^4 \left(56 M_{+}^4-85 M_{-}^4\right)  + M_X^2 \left(96 M_{+}^2 M_{-}^4 - 86 M_{+}^6\right)\nn\\
&  - 35 M_{+}^4 M_{-}^4 + 38
    M_{-}^8 \Big) \nn\\
  &  - \modulus{\Fe{1}{1}}^2 \left( \left(M_X^2 - 2
  M_{+}^2\right)^2 - M_{-}^4 \right) \Big(4 M_X^6 M_{+}^2\nn\\
&  - M_X^4
  \left( 5M_{+}^4 + M_{-}^4 \right) \nn\\
  & + 12 M_X^2 M_{+}^2
  \left( M_{+}^4 - M_{-}^4 \right) + 3 M_{+}^4 M_{-}^4-M_{-}^8
  \Big)\bigg),\\
  \Utwo{1} &= -\frac{8 M_1^2 M_2^2 M_{-}^4}{D_2^2}
  \modulus{\Fe{1}{1}}^2,\\
  \Vone{1} &= -\frac{9 \pi ^2 \eta^2 M_X M_1 M_2 Y}{2 \sqrt{2} D_1
    D_2} \left(M_X^4 - 2 M_X^2 M_{+}^2 - M_{-}^4\right)\nn\\ 
    &\qquad \times\; \Im\left(
  \Fe{1}{1} \Fost{1}{1} \right),\\
  \Vtwo{1} &= 0,\\
  \Tone{2} &= -\frac{3 \eta}{2 M_{+}^2 \mu^2 \nu^2} \bigg( 2 M_X
  M_{+}^3 \nu^2 \, \Re\left(\Fe{2}{2} \Fost{2}{2}\right) \nn\\
 &+ M_1^2 \mu^2
  \nu^2 \, \Re\left(\Fe{3}{2} \Fost{1}{2}\right) + M_2 M_{-}^2  \nn\\
& \times \bigg(\sqrt{3} M_1 \nu^2 \,
  \Re\left(\Fe{3}{2} \Fost{2}{2}\right)+ M_1 \mu^2 \,
  \Re\left(\Fe{4}{2} \Fost{1}{2}\right) \nn\\
   &+ \sqrt{3} M_2 M_{-}^2 \,
  \Re\left(\Fe{4}{2} \Fost{2}{2}\right) \bigg) \bigg),\\
  \Tonep{2} &= \frac{3 \eta}{2 M_{+}^2 \mu^2 \nu^2} \bigg( 2 M_X
  M_{+}^3 \nu^2 \, \Re\left(\Fe{2}{2} \Fost{2}{2}\right) \nn\\
&+ M_2^2 \mu^2
  \nu^2 \, \Re\left(\Fe{3}{2} \Fost{1}{2}\right) + M_1 M_{-}^2 \nn\\
  &  \times \left(-\sqrt{3} M_2 \nu^2 \,
\Re\left(\Fe{3}{2} \Fost{2}{2}\right) - M_2 \mu^2 \,
\Re\left(\Fe{4}{2} \Fost{1}{2}\right)\right) \nn\\
&+ \left(\sqrt{3} M_1^2 M_{-}^4 \,
\Re\left(\Fe{4}{2} \Fost{2}{2}\right)\right) \bigg),\\
  \Ttwo{2} &= \frac{1}{4}\bigg(-2 \modulus{\Fe{1}{2}}^2
  +\modulus{\Fe{2}{2}}^2 + \left(\modulus{\Fe{3}{2}}^2 +
  \modulus{\Fo{1}{2}}^2 \right) \left(\frac{M_1^2-2
    M_2^2}{M_{+}^2}\right) \nn\\
&+ \modulus{\Fe{4}{2}}^2 \left(2 M_X^2
  \frac{M_{+}^2}{\nu^4} + \frac{M_{-}^4}{M_{+}^2 \nu^4} \left(M_2^2-2
  M_1^2\right) \right) \nn \\
  &  +\modulus{\Fo{2}{2}}^2 \left(4 M_X^2
\frac{M_{+}^2}{\mu^4} + 3 \frac{M_{-}^4}{M_{+}^2 \mu^4} \left(M_2^2-2
M_1^2\right) \right) \nn\\
&+ \frac{6 M_1 M_2 M_{-}^2}{M_{+}^2 \mu^2 \nu^2}
\left(\mu^2 \, \Re\left(\Fe{3}{2} \Fest{4}{2}\right)+\sqrt{3} \nu^2 \,
\Re\left(\Fo{1}{2} \Fost{2}{2}\right)\right) \bigg),\\
  \Ttwop{2} &= \frac{1}{4} \bigg( -2 \modulus{\Fe{1}{2}}^2 +
  \modulus{\Fe{2}{2}}^2 + \left( \modulus{\Fe{3}{2}}^2 +
  \modulus{\Fo{1}{2}}^2 \right) \left(\frac{M_2^2-2
    M_1^2}{M_{+}^2}\right) \nn\\
&+ \modulus{\Fe{4}{2}}^2 \left(2 M_X^2
  \frac{M_{+}^2}{\nu^4} + \frac{M_{-}^4}{M_{+}^2 \nu^4} \left(M_1^2-2
  M_2^2\right)\right) \nn \\
  & + \modulus{\Fo{2}{2}}^2 \left(4 M_X^2
\frac{M_{+}^2}{\mu^4} +3 \frac{M_{-}^4}{M_{+}^2 \mu^4} \left(M_1^2-2
M_2^2\right)\right) \nn\\
  & -  \frac{6 M_1 M_2 M_{-}^2}{M_{+}^2 \mu^2 \nu^2}
\left(\mu^2 \, \Re\left(\Fe{3}{2} \Fest{4}{2}\right)+\sqrt{3} \nu^2 \,
\Re\left(\Fo{1}{2} \Fost{2}{2}\right)\right) \bigg),\\
  \Uone{2} &= \frac{9 \pi ^2 \eta ^2}{64 M_{+}^2 \mu^4 \nu^4}
  \bigg(\sqrt{2} M_{+}^2 \mu^4 \nu^4 \, \Re\left(\Fe{1}{2}\Fest{2}{2}\right)\nn\\
 &-M_{-}^4 \mu^4 \nu^2 \, \Re\left(\Fe{3}{2}
  \Fest{4}{2}\right)+\modulus{\Fe{3}{2}}^2 M_1 M_2 \mu^4 \nu^4 \nn\\
  &  -\modulus{\Fe{4}{2}}^2 M_1 M_2 M_{-}^4 \mu^4 + \sqrt{3}
M_{-}^4 \mu^2 \nu^4 \, \Re\left(\Fo{1}{2}\Fost{2}{2}\right)\nn\\
&-\modulus{\Fo{1}{2}}^2 M_1 M_2 \mu^4 \nu^4+3
\modulus{\Fo{2}{2}}^2 M_1 M_2 M_{-}^4 \nu^4\bigg),\\
  \Utwo{2} &= \frac{1}{4}\modulus{\Fe{2}{2}}^2 - \frac{M_X^2
    M_{+}^2}{\mu^4} \modulus{\Fo{2}{2}}^2,\\
  \Vone{2} &=\frac{9 \pi ^2 \eta ^2}{64 M_{+}^2 \mu^2 \nu^2} \bigg(2
  \sqrt{2} M_X M_{+}^3 \nu^2 \, \Im\left(\Fe{1}{2}\Fost{2}{2}\right)\nn\\
&+2 M_1 M_2 \mu^2 \nu^2 \, \Im\left(\Fe{3}{2}
  \Fost{1}{2}\right)- \sqrt{3} M_{-}^4 \nu^2 \,\Im\left(\Fe{3}{2} \Fost{2}{2}\right)\nn\\
 &- M_{-}^4 \mu^2 \,
\Im\left(\Fe{4}{2} \Fost{1}{2}\right) - 2 \sqrt{3} M_{-}^4 M_1 M_2 \,
\Im\left(\Fe{4}{2} \Fost{2}{2}\right) \bigg),\\
  \Vtwo{2} &= M_X \frac{M_{+}}{\mu^2} \, \Im\left(\Fe{2}{2}
  \Fost{2}{2}\right).
\end{align}
Here $\eta$ is defined as 
\begin{equation}
 \eta = \frac{2 \; v_\ell \; a_\ell}{v_\ell^2 + a_\ell^2},
\end{equation}
with $v_\ell = 2 I_{3\ell} - 4 e_\ell \; \sin^2\theta_W$ and $a_\ell = 2
I_{3\ell}$.  In the present case $v_\ell = -1 + 4 \sin^2\theta_W$ and $a_\ell
=-1$.  Substituting the value $\sin^2\theta_W = 0.231$, we get $\eta = 0.151$.


\begin{thebibliography}{99}
\bibitem{ref:Higher-spin} In this work we do not consider the spin-3
possibility. For the spin-3 case there would be two parity-even
transversity amplitudes and one parity-odd transversity amplitude.
Therefore, it can, in general, be distinguished from the spin-1 case
for which we have only two transversity amplitudes, one parity-even
and the other parity-odd. The spin-4 and higher spin cases are
similarly not considered, because all higher even spin cases have six
transversity amplitudes (the same number as the spin-2 case) and all
higher odd spin cases have three transversity amplitudes (the same
number as the spin-3 case).
\end{thebibliography}
\end{document}